\documentclass[12pt]{article}
\usepackage{showkeys}

\usepackage{mathrsfs}

\usepackage{amsmath,amssymb}
\usepackage{amsmath}
   \usepackage{amscd}
   \usepackage{amsfonts}
   \usepackage{amssymb}
   \usepackage{amsthm}
   \usepackage{epsfig}

\def\be{\begin{equation}}
\def\ee{\end{equation}}
\def\ba{\begin{eqnarray}}
\def\ea{\end{eqnarray}}
\def\str{\mbox{STr}}

\def\ket[#1]{\left|#1\right>}
\def\bra[#1]{\left<#1\right|}

\def\Zop{\bbbz}
\def\bbbz {{\sf Z\!\!Z}}

\def\R{{\bf R}}
\def\dg{\dagger}
\def\a{\alpha}
\def\at{{\tilde\a}}
\def\b{\beta}

\def\p{\partial}

\def\psib{{\bar\psi}}

\def\diag{\mbox{diag}}
\def\Db{{\bar D}}

\def\tX{{\tilde X}}
\def\tY{{\tilde Y}}
\def\tU{{\tilde U}}
\def\tV{{\tilde V}}

\def\tu{{\tilde u}}
\def\tv{{\tilde v}}
\def\tgamma{{\tilde\gamma}}
\def\trho{{\tilde\rho}}
\def\etab{{\bar\eta}}
\def\psib{{\bar\psi}}
\def\Psib{{\bar\Psi}}
\def\thetat{{\tilde\theta}}
\def\xit{{\tilde\xi}}
\def\thetab{{\bar\theta}}
\def\xib{{\bar\xi}}
\def\thetatb{{\bar{\tilde\theta}}}
\def\xitb{{\bar{\tilde\xi}}}

\def\Tr{{\rm  Tr}}
\def\tr{{\rm  tr}}
\def\STr{{\rm  STr}}
\def\Str{{\rm  Str}}

\def\ket[#1]{\left|#1\right>}
\def\bra[#1]{\left<#1\right|}

\def\ie{{\it i.e.}}

\parskip=\medskipamount

\input epsf
\usepackage{epsfig,color,latexsym}

\renewcommand{\theequation}{\thesection.\arabic{equation}}

\begin{document}
\thispagestyle{empty}
\def\thefootnote{\fnsymbol{footnote}}\begin{flushright} 
hep-th/yymmnnn\\ 
MIT-CTP-nnnn\\
Imperial/TP/2-07/nn
\end{flushright}\vskip 0.5cm\begin{center}
\LARGE{\bf  Landau-Lifshitz sigma-models, fermions
and the AdS/CFT correspondence}
\end{center}\vskip 0.8cm\begin{center}{\large 
B. Stefa\'nski, jr.$^{1,2}$
%\footnote{E-mail address: {\tt bogdans@lns.mit.edu}}
}
\vskip 0.2cm{\it $^1$ Center for Theoretical Physics \\ Laboratory for Nuclear Science, 
\\ Massachusetts Institute of Technology \\  Cambridge, MA  02139, USA}
\vskip 0.2cm{\it $^2$ Theoretical Physics Group,  Blackett  Laboratory, \\
Imperial College,\\ London SW7 2BZ, U.K.}
\end{center}
\vskip 1.0cm
\begin{abstract}\noindent
We define Landau-Lifshitz sigma models on general coset space $G/H$, with $H$ a maximal stability sub-group of $G$. These are 
non-relativistic models that have $G$-valued N\"other charges, local $H$ invariance and are classically integrable. Using this 
definition, we construct the $PSU(2,2|4)/PS(U(2|2)^2)$ Landau-Lifshitz sigma-model. This sigma model describes the thermodynamic 
limit of the spin-chain Hamiltonian obtained from the complete one-loop dilatation operator of the $N=4$ super Yang-Mills (SYM) 
theory. In the second part of the paper, we identify a number of consistent truncations of the Type IIB Green-Schwarz action on 
$AdS_5\times S^5$ whose field content consists of two real bosons and 4,8 or 16 real fermions. We show that $\kappa$-symmetry acts 
trivially in these sub-sectors. In the context of the large spin limit of the AdS/CFT correspondence, we map the Lagrangians of 
these sub-sectors to corresponding truncations of the $PSU(2,2|4)/PS(U(2|2)^2)$ Landau-Lifshitz sigma-model.
\end{abstract}

\vfill

\setcounter{footnote}{0}
\def\thefootnote{\arabic{footnote}}
\newpage

\renewcommand{\theequation}{\thesection.\arabic{equation}}
%%%%%%%%%%%%%%%%%%%%%%%%%%%%%%%%%%%%%%%%%%%%%%%%%%%%%%%%%%%%%%%%%%%%%%%%%

\section{Introduction}\label{sec1}
\setcounter{equation}{0}

The gauge/string correspondence~\cite{adscft} provides an amazing connection between quantum gauge and gravity
theories. The correspondence is best understood in the case of the maximally supersymmetric dual pair of
${\cal N}=4$ $SU(N)$ super-Yang-Mills (SYM)  gauge theory and Type IIB string theory on $AdS_5\times S^5$.
Recent progress in understanding this duality has come from investigations of states in the dual theories with
large charges~\cite{bmn,gkp2,ft}. In these large-charge limits (LCLs) it is possible to test the duality in
sectors where quantities are not protected by supersymmetry. Typically, one compares the energy of some
semi-classical string state with large charges (labelled schematically $J$) to the anomalous dimensions of the
corresponding operator in the dual gauge theory, using $1/J$ as an expansion parameter which supresses quantum
corrections. A crucial ingredient, which made such comparisons possible, was the observation that computing
anomalous dimensions in the ${\cal N}=4$ SYM gauge theory is equivalent to finding the energy eigenvalues of
certain integrable spin-chains~\cite{andim} (following the earlier work on more generic gauge
theories~\cite{oandim}). At the same time the classical Green-Schwarz (GS) action for the Type IIB string
theory on $AdS_5\times S^5$ was shown to be integrable~\cite{bpr}. The presence of integrable structures has 
led to an extensive use of Bethe ansatz-type techniques
to investigate the gauge/string duality~\cite{ba}.  In particular,
impressive results for matching the world-sheet S-matrix of the GS string sigma-model with the corresponding
S-matrix of the spin-chain have been obtained~\cite{smatrix}.

The matching of anomalous dimensions of gauge theory operators with the energies of semi-classical string 
states was shown to work up to and including two loops in the 't Hooft coupling $\lambda$. At three loops it 
was shown that the string and gauge theory results differ. As has been noted many times in the literature, 
this result should not be interpretted as a falsification of the gauge/string correspondence conjecture. 
Indeed, while the (perturbative) gauge theory computatons are done at small values in $\lambda$, they are 
compared to dual string theory energies which are computed at large values of $\lambda$ and as such are not 
necessarily comparable. It has then been a fortunate coincidence that the one- and two-loop results do match. 

This match was first established in a number of particular semi-classical string solutions and
corresponding single-trace operators~\cite{ft}. Later it was shown that, to leading order in the LCL, for
some bosonic sub-sectors the string action reduced to a generalised Landau-Lifshitz (LL) sigma model, which
also could be obtained as a thermodynamic limit of the corresponding spin-chain~\cite{k1,k2,hl1,st1,k3} (see
also~\cite{mikh}). In this way, by matching Lagrangians on both sides one can establish that energies of a
wide class of string solutions do indeed match with the corresponding anomalous dimensions of gauge theory
operators without having to compute these on a case-by-case basis. 

A natural extension of this programme is to
match, to leading order, the LCL of the full GS action of Type IIB string theory on $AdS_5\times S^5$ to the
thermodynamic limit of the spin-chain corresponding to the dilatation operator for the full ${\cal N}=4$ SYM
gauge theory; including fermions on both sides of the map is interesting given the different way in which they
enter the respective actions. On the spin-chain side fermions are on equal footing to
bosons~\cite{st2,hl2} - the LL equation, which describes the thermodynamic limit of the system, relates to a
super-coset manifold when fermions are included, as opposed to a coset manifold when there are no fermions. In
particular, both fermions and bosons satisfy equations which are first order in $\tau$ and second order in
$\sigma$. On the other hand, fermions in the GS action possess $\kappa$-symmetry~\cite{ws,gs,hm,mt} and their
equations of motion are first order both in $\tau$ and $\sigma$.  Previous progress on this question was able
to match string and spin chain actions in a LCL up to quadratic level in fermions~\cite{mikh,hl2,st2}.  
Roughly speaking, on the string side, $\kappa$-gauge fixed equations of motion for fermions typically come as
$2n$ first order equations. From these one obtains $n$ second-order equations for $n$ by 'integrating out'
half of the fermions. Taking a non-relativistic limit on the worldsheet one ends up with equations which are
first order in $\tau$ and second order in $\sigma$ which can be matched with the corresponding LL equations
obtained from the spin chain side. Matching the terms quartic and higher in the fermions had so far not been 
achieved, though it is expected that this should be possible given the results of~\cite{fullalgcurve}. 
However, finding a suitable $\kappa$-gauge in which this matching could be done in a natural way remained an
obstacle. Below we propose a $\kappa$ gauge which appears to be natural from the point of view of the 
dual spin-chain and allows for a matching of higher order fermionic terms in the dual Lagrangians. 

In this paper we first present a compact way of writing LL sigma models for quite general (super-)cosets
$G/H$; in particular we write down the full $PSU(2,2|4)/PS(U(2|2)^2)$ LL sigma model which arrises as the
thermodynamic limit of the one-loop dilatation operator for the full ${\cal N}=4$ SYM theory. This generalises
earlier work by~\cite{ran}, and allows one to write down LL-type actions without having to go through the
coherent-state~\cite{pere} thermodynamic limit of the spin chain. We then identify a number of sub-sectors of the
classical GS action~\footnote{By a sub-sector we mean that the classical equations of motion for the full GS
superstring on $AdS_5\times S^5$ admit a truncation in which all other fields are set to zero in a manner
which is consistent with their equations of motion. This is quite familiar in two cases: $(i)$ when one sets
all fermions in the GS action to zero and, (ii) when one further restricts the bosons to lie on some
$AdS_p\times S^q$ sub-space ($1\ge p\,,\,,q \ge 5$).} all of which have two real bosonic degrees of freedom and a
larger number of fermionic degrees of freedom (specifically 4,8 and 16 real fermionic d.o.f.s~\footnote{The 4
fermion model was previously postulated to be a sub-sector of the classical GS action in~\cite{aaf} and
represents a starting point for our analysis.}). Finally, we define a LCL in which the GS actions for these
fermionic sub-sectors reduce to corresponding LL actions. In this way we match the complete Lagrangians for
these sub-sectors and not just the terms quadratic in fermions. Since the largest of these sectors contains the 
maximal number of fermions (sixteen) for a $\kappa$-fixed GS action the LCL matching to a LL model gives a clear 
indication of what the natural $\kappa$-gauge is from the point of view of the dual spin-chain.

The fermionic sub-sectors of the GS action that we find are quite interesting in themselves because on-shell
$\kappa$-symmetry acts trivially on them - in particular the sub-sector containing 16 fermionic degrees of
freedom contains the same number of fermions as the $\kappa$-fixed GS superstring on $AdS_5\times S^5$. Since
$\kappa$-symmetry acts trivially in this case one cannot use it to eliminate half of the fermions as one does
in more conventional GS actions. Further, these fermionic sub-sectors naturally inherit the classical
integrability of the full GS superstring on $AdS_5\times S^5$ found in~\cite{bpr}.  Integrating out the metric
and the two bosonic degrees of freedom one then arrives at a new class of integrable differential equations
for fermions only. 

This paper is organised as follows. In section~\ref{sec2} we give a prescription for constructing a LL sigma
model on a general coset $G/H$. We also present a number of explicit examples of LL sigma models most relevant
to the gauge/string correspondence there and in Appendix~\ref{appe}. In section~\ref{sec3} we identify the 
fermionic sub-sectors of the GS
superstring on $AdS_5\times S^5$. In section~\ref{sec4} we define a LCL in which the GS action of the
fermionic sub-sectors reduces, to leading order in $J$, to the LL sigma models for the corresponding
gauge-theory fermionic sub-sectors. Since the GS action for the four fermion subsector is quadratic in the 
remaining appendices to this paper we present a more detailed discussion of it including a light-cone 
quantisation in 
Appendix~\ref{appa}, a discussion of its conformal invariance in Appendix~\ref{appb} and a T-dual form of the action
in Appendix~\ref{appd}.

\section{Landau-Lifshitz sigma models}\label{sec2}
\setcounter{equation}{0}

In this section we construct the Lagrangian for a Landau-Lifshitz (LL) sigma model on a coset
$G/H$.~\footnote{For earlier work on this see~\cite{ran}.} The Lagrangian will typically be first
(second)  order in the worldsheet time (space) coordinate, and so is non-relativistic on the worldsheet. We
refer to such models as LL sigma models because in the case of $G/H=SU(2)/U(1)$ the equations of motion reduce
to the usual LL equation
\be
\p_\tau n_i = \varepsilon_{ijk}n_j\p_\sigma^2n_k\,,\qquad \mbox{where}\qquad
n_in_i=1\,.
\ee
                                                                                
\noindent The construction of LL Lagrangians is closely related to coherent
states $\ket[\omega,\Lambda]$. Recall~\footnote{For a detailed exposition of coherent
states see~\cite{pere}; a brief summary, using the same notation as 
in this paper, is also
presented in Appendix A of~\cite{st1}.} that to construct a coherent 
state $\ket[\omega\,,\Lambda]$ we
need to specify a unitary irreducible representation $\Lambda$ of $G$ acting
on a Hilbert space $V_\Lambda$ and a vacuum state $\ket[0]$ on which $H$ is a maximal stability
sub-group, in other words for any $h\in H$
\be
\Lambda(h)\ket[0]=e^{i\phi(h)}\ket[0]\,,\label{vacuumphase}
\ee
with $\phi(h)\in\R$. Given such a representation $\Lambda$ and
state $\ket[0]$ we define the operator $\Omega$ as
\be
\Omega\equiv \ket[0]\bra[0]\,.\label{Omega}
\ee
The LL sigma model Lagrangian on $G/H$ is defined as
\be
{\cal L}_{\mbox{\scriptsize LL G/H}}={\cal L}^{\mbox{\scriptsize
WZ}}_{\mbox{\scriptsize LL G/H}}
+{\cal L}^{\mbox{\scriptsize kin}}_{\mbox{\scriptsize LL G/H}}
\label{LLsigmamodel}
\ee
where
\ba
{\cal L}^{\mbox{\scriptsize WZ}}_{\mbox{\scriptsize LL G/H}}&=&
-i\Tr\left(\Omega g^\dg\p_\tau g\right)\,,
\label{LLLWZ}
\\
{\cal L}^{\mbox{\scriptsize kin}}_{\mbox{\scriptsize LL G/H}}&=&
\Tr\left(g^\dg D_\sigma gg^\dg D_\sigma g\right)\,.
\ea
Above, $g^\dg D_\sigma g\equiv g^\dg\p_\sigma g-g^\dg\p_\sigma g|_H$ is just
the standard $H$-covariant current. It is then clear that 
${\cal L}^{\mbox{\scriptsize kin}}_{\mbox{\scriptsize LL G/H}}$ 
is invariant under gauge transformations
\be
g\rightarrow gh\,,
\ee
for any $h=h(\tau\,,\,\sigma)\in H$. We may also show that the same
is true of 
${\cal L}^{\mbox{\scriptsize WZ}}_{\mbox{\scriptsize LL G/H}}$. To see this 
note that the gauge variation of 
${\cal L}^{\mbox{\scriptsize WZ}}_{\mbox{\scriptsize LL G/H}}$, using
equation~(\ref{vacuumphase}), is given by
\be
\delta_H {\cal L}^{\mbox{\scriptsize WZ}}_{\mbox{\scriptsize LL G/H}} =
e^{-i\phi(h)}\bra[0]\p_\tau h\ket[0]=e^{-i\phi(h)}\p_\tau(\bra[0]
h\ket[0])=i\p_\tau\phi(h)\,.
\ee
This in turn is a total derivative; and so the full action is invariant under local right $H$ action. The 
Lagrangian also has a global $G$ symmetry 
\be
g\rightarrow g_0 g\,,
\ee
for any $g_0\in G$ with $\p_\tau g_0=\p_\sigma g_0 =0$, and the corresponding N\"other current is given by
\be
(j_\tau\,,\,j_\sigma)=(g\Omega g^\dg,2iD_\sigma gg^\dg)\,.
\ee

In~\cite{st1,st2} LL actions were written down in terms of Lie algebra matrices denoted typically by $N$. To make 
contact with the present notation we note that~\footnote{The following equation is due to Charles Young.}
\be
N\equiv g\Omega g^\dagger -\frac{1}{n}{\bf I}_n\,,
\ee
where the second term on the right hand side is included since $N$ is traceless.
Finally,
let us note that these LL sigma models admit a Lax pair representation and as a result are integrable. This is most 
easily seen in terms of the matrix $N$ for which the equations of motion are the LL matrix equation 
\be
\p_\tau N =\frac{i}{2}\left[N\,,\,\p_\sigma^2 N\right\}\,.
\ee
This is equivalent to the zero-curvature condition on the following Lax pair
\ba
{\cal L}&\longrightarrow& \p_\sigma -\frac{iN}{4\pi x}\,,\\
{\cal M}&\longrightarrow& \p_\tau -\frac{iN}{4\pi^2 x^2}-
\frac{[N,\p_\sigma N\}}{8\pi x}\,,
\ea
where $\left[\cdot\,,\,\cdot\right\}$ is the (super)-commutator. In the remainder of this section we construct a
number of explicit examples of LL sigma models. Further examples of interest in the gauge/string correspondence are
relagated to Appendix~\ref{appe}. The reader who is not interested in the details of these examples should skip the
remainder of this section.

\subsection{The $U(1|1)/U(1)^2$ model}

This is one of the simplest LL sigma 
models,~\footnote{There is also the equally simple bosonic 
U(1) LL sigma model.} in that the Lagrangian is quadratic
\be
{\cal L}_{\mbox{\scriptsize LL $U(1|1)/U(1)^2$
}}=i\psib\p_\tau\psi+\p_\sigma\psib\p_\sigma\psi\,,
\label{LLu11}
\ee
with $\psi$ a complex Grassmann-odd field and $\psib$ its complex conjugate. Notice that this
result can be obtained using the explicit $2\times 2$ supermatrix representation of $U(1|1)$, 
with the vacuum state $\ket[0]$ being the super-vector $(0,1)$.

\subsection{The $SU(3)/S(U(2)\times U(1)$ model}
                                                                                                          
Before proceeding to our main example - the $PSU(2,2|4)$ model - in this subsection we show how the
above formal prescription applies to the well known $SU(3)$ Landau-Lifshitz model~\cite{st1,hl1}. Recall
that the Lagrangian for this is
\be
{\cal L}_{\mbox{\scriptsize $SU(3)/S(U(2)\times U(1))$}}=
-iU^i\p_\tau U_i-\frac{1}{2}|D_\sigma U_i|^2+\Lambda(U_iU^i-1)\,,\label{gtsu3}
\ee
where
\be
D_\mu U_i\equiv\p_\mu -iC_\mu \,,\qquad C_\mu=-iU^i\p_\mu U_i\,,
\ee
for $\mu=\tau,\,\sigma$ and $U^i\equiv U_i^*$. To show that we
can obtain this from our general expression~(\ref{LLsigmamodel}) we
write elements of the group $SU(3)$ as $3\times 3$ matrix $g$,
split into a $3\times 2$ matrix $X$ and a vector $Y$
\be
g=\left(X,Y\right)\,,
\ee
and because $g$ is in $SU(3)$ (\ie{} $g^\dagger g=1$) we have
\ba
X^\dagger X ={\bf 1}_2\,,\qquad Y^\dagger Y={\bf 1}_1\,,&\qquad&
X^\dagger Y=0\,,\qquad Y^\dagger X=0\,,\\
XX^\dagger+YY^\dagger &\!\!\!\!\!\!\!\!=&\!\!\!\!\!\!\!\!{\bf 1}_3\,.
\ea
The kinetic part of the Lagrangian~(\ref{LLsigmamodel}) is then given by
\ba
{\cal L}_{\mbox{\scriptsize kin $SU(3)/S(U(2)\times U(1))$}}
&=&\frac{1}{4}\Tr\left((g^{-1}D_1g)(g^{-1}D_1g)\right)
\nonumber \\
&=&\frac{1}{4}\Tr\left[
\left(\begin{array}{cc}X^\dg D_1 X & X^\dg D_1 Y \\ Y^\dg D_1 X & Y^\dg D_1 Y \end{array}\right)^2\right]
=\frac{1}{4}\Tr\left[
\left(\begin{array}{cc}0 & X^\dg \p_1 Y \\ Y^\dg \p_1 X & 0 \end{array}\right)^2\right]
\nonumber \\
&=&\frac{1}{2}\Tr\left[X^\dg\p_1 Y Y^\dg\p_1X\right]
\nonumber \\
&=&-\frac{1}{2}\Tr\left[\p_1X^\dg Y Y^\dg\p_1X\right]=-\frac{1}{2}\Tr\left[\p_1Y^\dg X X^\dg\p_1Y\right]
\nonumber \\&=&\frac{1}{2}
-\Tr\left[\p_1X^\dg(1-XX^\dg)\p_1X\right]=-\frac{1}{2}\p_1Y^i(\delta_i^j-Y_iY^j)\p_1Y_j
\nonumber \\
&=&-\frac{1}{2}\Tr\left[\Db_1X^\dg D_1X\right]=-\frac{1}{2}\Db_1Y^i D_1Y_i\,.
\ea
The final expression is the same as the kinetic term of the usual $SU(3)$ Landau-Lifshitz
Lagrangian~(\ref{gtsu3}) upon identifying $Y_i$ with $U_i$ (above $Y^i\equiv Y^\dg$).
Above, we have defined
\ba
D_1Y_i&=&\p_1Y_i-Y_iY^j\p_1Y_j\,,\qquad \Db_1 Y^i\equiv (D_1Y_i)^\dagger  \\
D_1X&=&\p_1X-XX^\dagger\p_1X\,,\qquad \Db_1 X\equiv (D_1X)^\dagger\,.
\ea
The WZ term of the Lagrangian is given by equation~(\ref{LLLWZ}) and can be written as
\be
{\cal L}_{\mbox{\scriptsize WZ $SU(3)/S(U(2)\times U(1))$}}=i\,
\Tr(X^\dg\p_0 X)=-iY^i\p_0Y_i\,.
\ee
This follows from the fact that $g^{-1}\p_0g$ is traceless and so
\be
\Tr(X^\dg\p_0 X)=-Y^i\p_0Y_i\,.
\ee
Upon identifying $Y_i$ with $U_i$, the WZ term above
is the same as the usual $SU(3)$ Landau-Lifshitz one~(\ref{gtsu3}). Notice
that we have also given an alternate parametrisation of the $SU(3)$ Landau-Lifshitz
model in terms of $X$
\be
{\cal L}_{\mbox{\scriptsize $SU(3)/S(U(2)\times U(1))$}}=
i\,\Tr(X^\dg\p_0 X)-\frac{1}{2}\Tr\left[\Db_1X^\dg D_1X\right]+\Lambda(X^\dg X-1_2)\,,
\ee
which has an explicit $SU(2)$ gauge invariance.
                                                                                                          
Finally, out of $X$ and $Y$ we may define a matrix which takes values in the $SU(3)$
Lie algebra
\be
N^i{}_j=3Y^iY_j-\delta^i_j=-3X_{ja}X^{ai}+2\delta^i_j\,,
\ee
where $a=1,2$. This matrix is however, not a general $SU(3)$ matrix but rather satisfies the identity
\be
N^2=N+2\,.
\ee
In terms of $N$ the equations of motion take the form of the matrix Landau-Lifshitz equation
\be
\p_0 N=-\frac{i}{9}[N\,,\p_1^2N]\,.
\ee
These are equivalent to the consistency of the following linear problem
\ba
{\cal L}{\cal \psi}&=&\left[\p_\sigma-\frac{i}{4\pi x}N\right]{\cal\psi}=0\,,\\
{\cal M}{\cal \psi}&=&\left[\p_\tau-\frac{i}{4\pi^2x^2}N-\frac{b}{4\pi x}
%(Y\Db_\sigma Y^\dg-D_\sigma YY^\dg)
[N,\p_1N]
\right]{\cal\psi}=0\,.
\ea

\subsection{The $SU(2,2|4)/S(U(2|2)\times U(2|2))$ model}
                                                                                                          
In this sub-section we present an explicit Lagrangian for the complete $PSU(2,2|4)$ Landau-Lifshitz
sigma model Lagrangian following the general discussion at the start of the present section. The
action we are interested in is the Landau Lifshitz model as defined in equation~(\ref{LLsigmamodel})
on the coset
\be
\frac{PSU(2,2|4)}{PS(U(2|2)\times U(2|2))}\,,
\ee
or on the coset
\be
\frac{SU(2,2|4)}{S(U(2|2)\times U(2|2))}\,,
\ee
both of which have 32 real components. The derivation is very similar to the $SU(3)$ Lagrangian
derived in the previous sub-section, and so we will simply state our results. A general group
element $g$ can be written as $(X,Y)$ where now $X$ ($Y$) is a $8\times 4$ supermatrix, with the
diagonal $4\times 4$ blocks bosonic (fermionic) and the off-diagonal $4\times 4$ blocks fermionic
(bosonic). The Lagrangian is then given by
\be
{\cal L}_{\mbox{\scriptsize LL
$PSU(2,2|4)/PS(U(2|2)\times U(2|2))$}}
=
i\STr(X^\dg\p_0X)-\frac{1}{2}\STr(\Db_1X^\dg D_1X)+\Lambda(X^\dg X-1)\,.
\ee
Note that there are 32 complex degrees of freedom in $X$, which the constraints reduce to 48 real
degrees of freedom. The action also has a local $U(2|2)$ gauge invariance, so in total the above
Lagrangian has 32 degrees of freedom - the same as the coset.
                                                                                                          
\noindent
In fact we may write $X$ as
\be
X=(\tU_a,\tV_a,U_a,V_a)\,,\qquad
X^\dg\equiv(\tU^a,\tV^a,U^a,V^a)\,,
\ee
where $a=1,\dots,8$, and
\ba
\tU^a\tU_a&=&-1\,,\qquad\!\!\!
\tV^a\tV_a=-1\,,\qquad
\tV^a\tU_a=0\,,\qquad
\tU^a\tV_a=0\,,\\
U^aU_a&=&1\,,\qquad
V^aV_a=1\,,\qquad
V^aU_a=0\,,\qquad
U^aV_a=0\,,\\
U^a\tU_a&=&0\,,\qquad
U^a\tV_a=0\,,\qquad
V^a\tU_a=0\,,\qquad
V^a\tV_a=0\,,\\
\tU^aU_a&=&0\,,\qquad
\tU^aV_a=0\,,\qquad
\tV^aU_a=0\,,\qquad
\tV^aV_a=0\,.
\ea
Above we have defined
\be
U^a=U_b^*C^{ba}\,,\qquad
V^a=V_b^*C^{ba}\,,\qquad
\tU^a=-\tU_b^*C^{ba}\,,\qquad
\tV^a=-\tV_b^*C^{ba}\,,
\ee
where $C^{ab}=\diag(-1,-1,1,1,1,1,1,1)$.
                                                                                                          
\noindent
The Lagrangian ~(\ref{psu224}) written in terms of $\tU_a,\tV_a,U_a,V_a$ is
\ba
{\cal L}_{\mbox{\scriptsize LL $PSU(2,2|4)/PS(U(2|2)^2)$}}
&=&
-i\tU^a\p_0\tU_a-i\tV^a\p_0\tV_a
-iU^a\p_0U_a-iV^a\p_0V_a
\nonumber \\
&&
-\frac{1}{2}\Bigl(
\p_1\tU^a\p_1\tU_a+\p_1\tV^a\p_1\tV_a
+\p_1U^a\p_1U_a+\p_1V^a\p_1V_a
\nonumber \\&&\,\,\,\,\,\,\,\,\,\,\,\,
-\tU^a\p_1\tU_a\tU^b\p_1\tU_b
-\tV^a\p_1\tV_a\tV^b\p_1\tV_b
\nonumber \\&&\,\,\,\,\,\,\,\,\,\,\,\,
+V^a\p_1V_aV^b\p_1V_b
+U^a\p_1U_aU^b\p_1U_b
\nonumber \\&&\,\,\,\,\,\,\,\,\,\,\,\,
+2V^a\p_1U_aU^b\p_1V_b
-2\tV^a\p_1\tU_a\tU^b\p_1\tV_b
+2\tU^a\p_1U_aU^b\p_1\tU_b
\nonumber \\&&\,\,\,\,\,\,\,\,\,\,\,\,
+2\tU^a\p_1V_aV^b\p_1\tU_b
+2\tV^a\p_1U_aU^b\p_1\tV_b
+2\tV^a\p_1V_aV^b\p_1\tV_b
\Bigr)\nonumber \\
\,.\label{psu224}
\ea
One can check explicitly that this action has local U(2$|$2) invariance
\be
(\tU_a,\tV_a,U_a,V_a)\rightarrow (\tU_a,\tV_a,U_a,V_a)U(\tau,\sigma)\,,
\ee
for $U$ a U(2$|$2) matrix.

\subsubsection{Subsectors of the the $SU(2,2|4)/S(U(2|2)\times U(2|2))$ model}
                                                                                                          
In the above Lagrangian we may set
\be
\tU_a=(1,0^7)\,,\qquad
\tV_a=(0,1,0^6)\,,\qquad
U_a=(0^2,U_3,\dots,U_8)\,,\qquad
V_a=(0^2,V_3,\dots,V_8)\,,
\ee
where
\be
U^aU_a=1\,,\qquad
V^aV_a=1\,,\qquad
V^aU_a=0\,,\qquad
U^aV_a=0\,.
\ee
The resulting Lagrangian is that of the SU(2$|$4) sector. If we further set
\be
0=U_3=U_4=V_3=V_4\,,
\ee
we can recover the SO(6) Lagrangian~(\cite{st1}). Details of this are
presented in Appendix~\ref{appa}. We may further consistently set
\be
0=U_8=V_3=V_4=V_5=V_6=V_7\,,\qquad V_8=1\,,
\ee
in which case we obtain the SU(2$|$3) Lagrangian~(\cite{st2}), with the identification
$(U_3,U_4)\equiv(\psi_1,\psi_2)$.
                                                                                                          
\noindent
We may instead set
\be
U_a=(0^7,1)\,,\qquad
V_a=(0^6,1,0)\,,\qquad
\tU_a=(U_1,\dots,U_6,0^2)\,,\qquad
\tV_a=(V_1,\dots,V_6,0^2)\,,
\ee
where
\be
\tU^a\tU_a=-1\,,\qquad
\tV^a\tV_a=-1\,,\qquad
\tV^a\tU_a=0\,,\qquad
\tU^a\tV_a=0\,.
\ee
The resulting Lagrangian is that of the SU(2,2$|$2) sector. If we further set
\be
0=U_3=U_4=V_3=V_4\,,
\ee
we recover the SO(2,4) Lagrangian, which is the Wick rotated version of
the $SO(6)$ Lagrangian~(\cite{st1}). In Appendix~\ref{appa} we write out this Lagrangian explicitly.
                                                                                                          
\noindent
A final interesting choice is to set
\be
U_a=(0^7,1)\,,\qquad
V_a=(0,V_2,\dots,V_7,0)\,,\qquad
\tU_a=(0,U_2,\dots,U_7,0)\,,\qquad
\tV_a=(1,0^7)\,,
\ee
where
\be
\tU^a\tU_a=-1\,,\qquad
V^aV_a=1\,,\qquad
V^a\tU_a=0\,,\qquad
\tU^aV_a=0\,.
\ee
The resulting Lagrangian is that of the SU(1,2$|$3) sector.
If we further set
\be
0=V_2=V_7=\tU_2=\tU_7\,,
\ee
we get the $SU(2|2)$ Lagrangian. In Appendix~\ref{appa} we write out this Lagrangian explicitly.

\section{Green-Schwarz actions and fake $\kappa$-symmetry}\label{sec3}
\setcounter{equation}{0}

in this section we construct GS sigma model actions whose field content are two real bosons and 
4,8 or 16 real fermions. These models all come from consistent truncations of the equations of motion for the 
full Type IIB GS action on $AdS_5\times S^5$. Just as any GS sigma model these fermionic actions have a
$\kappa$-symmetry. However, we show that for these models $\kappa$-symmetry is trivial on-shell. As a result 
one cannot use it to reduce the fermionic degrees of freedom of these models by fixing a $\kappa$-gauge as one 
does in more conventional GS actions.

Let us briefly recall the construction of the GS action on a super-coset 
$G/H$. We require that: (i) $H$ be bosonic and, (ii) $G$ admit a $\Zop_4$ automorphism 
that leaves $H$ invariant, acts by $-1$ on the remaining bosonic part of $G/H$, and by $\pm i$ on the 
fermionic part of $G/H$. The currents
$j_\mu=g^\dg\p_\mu g$ can then be decomposed as 
\be
j_\mu=j^{(0)}_\mu+j^{(1)}_\mu+j^{(2)}_\mu+j^{(3)}_\mu\,,\label{currz4dec}
\ee
where $j^{(k)}$ has eigenvalue $i^k$ under the $\Zop_4$ automorphism. In terms of these the GS action can be 
written as
\be
{\cal L}_{\mbox{\scriptsize GS $G/H$}}=\int d^2\sigma\,\,
\sqrt{-g}g^{\mu\nu}\Str(j^{(2)}_\mu j^{(2)}_\nu)
+\epsilon^{\mu\nu}\Str(j^{(1)}_\mu j^{(3)}_\nu)\,,\label{z4gs}
\ee
from which the equations of motion are
\ba
0&=&\p_\a(\sqrt{-g}g^{\a\b}j^{(2)}_\b)-\sqrt{-g}g^{\a\b}\left[j^{(0)}_\a,j^{(2)}_\b\right]
+\frac{1}{2}\epsilon^{\a\b}\left(\left[j^{(1)}_\a,j^{(1)}_\b\right]-\left[j^{(3)}_\a,j^{(3)}_\b\right]
\right)\,,\label{eom1}\\
0&=&\left(\sqrt{-g}g^{\a\b}+\epsilon^{\a\b}\right)\left[j^{(3)}_\a,j^{(2)}_\b\right]\,,\label{eom2}\\
0&=&\left(\sqrt{-g}g^{\a\b}-\epsilon^{\a\b}\right)\left[j^{(1)}_\a,j^{(2)}_\b\right]\,.\label{eom3}
\ea

\subsection{Fermionic GS actions}

Having briefly reviewed the general construction of GS actions on $G/H$ super-cosets, we now turn to the main 
focus of this section which is identifying GS actions with a large number of fermionic degrees of freedom, which are 
consistent truncations of the full $AdS_5\times S^5$ GS action. To do this consider the following
sequence of super-cosets
\be
\frac{U(1|1)\times U(1|1)}{U(1)\times U(1)}\subset
\frac{U(2|2)}{SU(2)\times SU(2)}\subset
\frac{PS(U(1,1|2)\times U(2|2))}{SU(1,1)\times SU(2)^3}\subset
\frac{PSU(2,2|4)}{SO(1,4)\times SO(5)}\,.
\ee
The $\subset$ symbols are valid both for the numerators and denominators and hence for the cosets as written 
above. Notice that the right-most of these cosets is just the usual Type IIB on $AdS_5\times S^5$ super-coset. 
Further, it is easy to convince onself that each of the cosets above admits a $\Zop_4$ automorphism which is 
compatible with the $\Zop_4$ automorphism of the Type IIB on $AdS_5\times S^5$ super-coset. The $\Zop_4$ 
automorphisms may be used to write down GS actions for each of these cosets. The fact that the cosets embed into 
each other as shown above in a manner compatible with the $\Zop_4$ automorphism implies that their GS actions can 
be thought of as coming from a consistent truncation of the GS action of any coset to the right of it in the 
above sequence. In particular this reasoning shows that the GS actions for $U(1|1)^2/U(1)^2$, $U(2|2)/SU(2)^2$ 
and $U(1,1|2)\times U(2|2)/(SU(1,1)\times SU(2)^3$ can all be thought of as coming from consistent truncations of 
the Type IIB GS action on $AdS_5\times S^5$.

\noindent Counting the number of bosonic and fermionic components of the three cosets $U(1|1)^2/U(1)^2$, 
$U(2|2)/SU(2)^2$ and $U(1,1|2)\times U(2|2)/(SU(1,1)\times SU(2)^3$ we see immediately that they each have 2 real 
bosonic components and, respectively, 4,8 and 16 real fermionic components - which is why we refer to these 
actions as fermionic GS actions. We might expect that some of the femrionic degrees of freedom could be 
eliminated from the GS actions by fixing $\kappa$-symmetry. In fact, it turns out that for these models 
$\kappa$-symmetry acts trivially on-shell and so cannot be used to eliminate some of the fermionic degrees of 
freedom. Indeed, the GS actions on the above-mentioned cosets do have 4,8 and 16 real fermionic degrees of 
freedom, respectively.

\noindent In the remainder of this sub-section we write down explicitly the GS actions for $U(2|2)/SU(2)^2$ and
$U(1|1)^2/U(1)^2$ and discuss their $\kappa$ and gauge transformations; the GS action for 
$U(1,1|2)\times U(2|2)/SU(1,1)\times SU(2)^3$ may also be written down in an analogous fashion but since we will not
need its explicit form later we refrain from writing it out in full.

\subsection{The GS action on $U(2|2)/SU(2)^2$}

The GS action on action on $U(2|2)/SU(2)^2$ can be written 
down in terms of the parametrisation of the $U(2|2)$ supergroup-valued matrix written as
\be\label{ads2s2param}
g=(X,Y;\tX,\tY)\,,
\ee
where $X$, $Y$ ($\tX$, $\tY$) are four-component super-vectors with the first (last) two entries Grassmann
even and the last (first) two entries Grassmann odd. Since the matrix $g$ is unitary we must have
\ba
1&=&X^\dagger X =Y^\dagger Y=\tX^\dagger \tX =\tY^\dagger \tY\,,\nonumber \\
0&=&X^\dagger Y=Y^\dagger X=X^\dagger \tX=\tX^\dagger X=X^\dagger \tY=\tY^\dagger X
\nonumber\\ &=&Y^\dagger \tX=\tX^\dagger Y=Y^\dagger \tY=\tY^\dagger Y
=\tX^\dagger \tY=\tY^\dagger \tX\,,\nonumber \\
1_{(2|2)}&=&XX^\dagger+YY^\dagger+ \tX\tX^\dagger+\tY\tY^\dagger \,,
\label{ads2s2const}
\ea
where the matrix $1_{(2|2)}$ is just the $4\times 4$ identity matrix. The $\Zop_4$ automorphism is given by
\be
\label{Z4autu22}
\Omega:\,M=\left(\begin{array}{cc} A& B \\ C& D \end{array}\right)\longrightarrow
\left(\begin{array}{cc}\sigma^2&0\\0&\sigma^2\end{array}\right)
\left(\begin{array}{cc} -A^T& C^T \\ -B^T& -D^T \end{array}\right)
\left(\begin{array}{cc}\sigma^2&0\\0&\sigma^2\end{array}\right)\,,
\ee
\noindent
which acts on the current as
\be
\Omega(j_\mu)=\left(\begin{array}{cccc}
-Y^\dg\p_\mu Y & X^\dg\p_\mu Y & -\tY^\dg\p_\mu Y &  \tX^\dg\p_\mu Y \\
Y^\dg\p_\mu X & X^\dg\p_\mu X & \tY^\dg\p_\mu X & -\tX^\dg\p_\mu X \\
Y^\dg\p_\mu \tY & -X^\dg\p_\mu \tY  & -\tY^\dg\p_\mu \tY & \tX^\dg\p_\mu \tY \\
-Y^\dg\p_\mu \tX & X^\dg\p_\mu \tX & \tY^\dg\p_\mu \tX & -\tX^\dg\p_\mu \tX 
\end{array}\right)\,.
\ee
The Green-Schwarz action then is
\ba
{\cal L}_{\mbox{\scriptsize GS } U(2|2)/(SU(2)\times SU(2))}&=&\frac{1}{2}\int d^2\sigma 
\,\,\sqrt{g}g^{\mu\nu}\left( (X^\dg\p_\mu X+Y^\dg\p_\mu Y)(X^\dg\p_\nu X+Y^\dg\p_\nu Y)\right.
\nonumber \\&&\qquad\qquad\,\,\,\,\,\,\,\,\,\,\,\,\,\,\,\,\,\left.
-(\tX^\dg\p_\mu \tX+\tY^\dg\p_\mu \tY)(\tX^\dg\p_\nu \tX+\tY^\dg\p_\nu \tY)
\right)\nonumber \\&&\qquad\,\,\,\,\,\,\,\,\,\,\,\,
+2i\epsilon^{\mu\nu}\left(X^\dg\p_\mu \tX Y^\dg\p_\nu\tY
+\tY^\dg\p_\mu Y\tX^\dg\p_\nu X\right.\nonumber \\&&\qquad\qquad\qquad\,\,\,\,\left.
-X^\dg \p_\mu\tY Y^\dg\p_\nu\tX
-\tX^\dg\p_\mu Y\tY^\dg\p_\nu X\right)\,.\label{u22gs}
\ea
One can easily check that this action has a local $SU(2)\times SU(2)$ invariance which acts
on the doublets $(X,Y)$ and $(\tX,\tY)$. The action also has $\kappa$-symmetry which acts 
on the fields as~\footnote{The $\kappa$-action below has the nice feature of acting as a local fermionic group action
by multiplication from the right. Such a representation was originally suggested in~\cite{mcarthur} and was
developed more fully for the $AdS_5\times S^5$ GS action in~\cite{glebnotes}; the formulas below 
are a simple extension of this latter construction to the coset at hand.}
\ba
\delta_\kappa X&=&-{\tilde X}({\bar\epsilon}_1+{\bar{\tilde\epsilon}}_1)
-{\tilde Y}({\bar\epsilon}_2+{\bar{\tilde\epsilon}}_2)\nonumber \\
\delta_\kappa Y&=&i{\tilde X}(\epsilon_2-{\tilde\epsilon}_2)
-i{\tilde Y}(\epsilon_1-{\tilde\epsilon}_1)\nonumber \\
\delta_\kappa \tX&=&X(\epsilon_1+{\tilde\epsilon}_1)
+iY({\bar\epsilon}_2-{\bar{\tilde\epsilon}}_2)\nonumber \\
\delta_\kappa \tY&=&X(\epsilon_2+{\tilde\epsilon}_2)
-iY({\bar\epsilon}_1-{\bar{\tilde\epsilon}}_1)\,,\label{ku22coord}
\ea
where
\ba
\epsilon_i&=&\Pi^{\a\b}_+(X^\dg\p_\a X+Y^\dg\p_\a Y+\tX^\dg\p_\a \tX+\tY^\dg\p_\a \tY)\kappa_{i\,,\,\b}\nonumber \\
{\tilde\epsilon}_i&=&\Pi^{\a\b}_-(X^\dg\p_\a X+Y^\dg\p_\a Y+\tX^\dg\p_\a \tX+\tY^\dg\p_\a \tY){\tilde\kappa}_{i\,,\,\b}\,,
\label{ku22eps}
\ea
for $i=1\,,\,2$ with $\kappa_{i\,,\,\b}$ and ${\tilde\kappa}_{i\,,\,\b}$ local Grassmann-odd parameters. The world-sheet
metric also varies as
\ba
\delta_\kappa(\sqrt{-g}g^{\a\b})&=&\Pi^{\a\gamma}_+\left(\kappa_{1\,,\,+}^\b(\tX^\dg\p_\gamma X-iY^\dg\p_\gamma\tY)
+\kappa_{2\,,\,+}^\b(\tY^\dg\p_\gamma X+iY^\dg\p_\gamma\tX)+\mbox{c.c.}\right)+\alpha\leftrightarrow\beta
\nonumber \\ &&
+\Pi^{\a\gamma}_-\left({\tilde\kappa}_{1\,,\,-}^\b(\tX^\dg\p_\gamma X+iY^\dg\p_\gamma\tY)
+{\tilde\kappa}_{2\,,\,-}^\b(\tY^\dg\p_\gamma X-iY^\dg\p_\gamma\tX)+\mbox{c.c.}\right)+\alpha\leftrightarrow\beta\,.
\nonumber \\
\label{ku22metric}
\ea
Notice that the above variation is consistent with the symmetries and the unimodularity of $\sqrt{-g}g^{\a\b}$ as long as 
\be
\kappa_i^\a=\Pi_+^{\a\b}\kappa_{i\,,\,\b}\,,\qquad
{\tilde\kappa}_i^\a=\Pi_-^{\a\b}{\tilde\kappa}_{i\,,\,\b}\,.
\ee
In the above formulas we have decomposed two-component vectors $v_\a$ as
\be
v^{\a}_\pm\equiv\Pi^{\a\b}_\pm v_\b\equiv\frac{1}{2}\left(\sqrt{-g}g^{\a\b}\pm\epsilon^{\a\b}\right)v_\b\,.
\ee

%We may derive the N\"other currents associated with the global right-action of $U\in U(2|2)$ for the
%action~(\ref{gsu22})
%\be
%g\rightarrow U g\,.
%\ee
%There are eight fermionic currents, which are given by
%\ba
%Q^i_{\mu\,j}&=&(X^\dg\p_\mu X+Y^\dg\p_\mu
%Y)(X^iX_j+Y^iY_j)+(\tX^\dg\p_\mu\tX+\tY^\dg\p_\mu\tY)(\tX^i\tX_j+\tY^i\tY_j)
%\nonumber \\& &
%+i\epsilon_\mu{}^\nu\left[
%Y^\dg\p_\nu\tY X^i\tX_j+X^\dg\p_\nu\tX Y^i\tY_j-\tX^\dg\p_\nu X \tY^i Y_j-\tY^\dg\p_\nu Y \tX^i X_j\right.
%\nonumber \\ && \left.\qquad\,\,\,\,\,
%-Y^\dg\p_\nu\tX X^i\tY_j-X^\dg\p_\nu\tY Y^i\tX_j+\tY^\dg\p_\nu X \tX^i Y_j+\tX^\dg\p_\nu Y \tY^i X_j\right]
%\,,\label{sucharges}
%\ea
%for $i=1,2$ and $j=3,4$, together with their complex conjugates. There are also eight bosonic currents, which
%are given by
%\ba
%J^i_{\mu\,j}&=&(X^\dg\p_\mu X+Y^\dg\p_\mu
%Y)(X^iX_j+Y^iY_j)-(\tX^\dg\p_\mu\tX+\tY^\dg\p_\mu\tY)(\tX^i\tY_j+\tY^i\tY_j)
%\nonumber \\& &
%-i\epsilon_\mu{}^\nu\left[
%Y^\dg\p_\nu\tY X^i\tX_j+X^\dg\p_\nu\tX Y^i\tY_j+\tX^\dg\p_\nu X \tY^i Y_j+\tY^\dg\p_\nu Y \tX^i X_j\right.
%\nonumber \\ && \left.\qquad\,\,\,\,\,
%-Y^\dg\p_\nu\tX X^i\tY_j-X^\dg\p_\nu\tY Y^i\tX_j-\tY^\dg\p_\nu X \tX^i Y_j-\tX^\dg\p_\nu Y \tY^i X_j\right]
%\,,\label{charges}
%\ea
%for $i,\,j=1,2$ and $i,\,j=3,4$, together with their complex conjugates.

\subsection{The GS action on $U(1|1)^2/U(1)^2$}

To obtain the GS action on $U(1|1)^2/U(1)^2$ we may simply set
\be
0=X_3=Y_4=\tX_1=\tY_2\,.
\ee
in the action~(\ref{u22gs}). This is because now the group element $g$ given in equation~(\ref{ads2s2param}) belongs to 
$U(1|1)^2\subset U(2|2)$; this truncation is also consistent with the $\Zop_4$ automorphism~(\ref{Z4autu22}).
As was argued at the start of this sub-section these facts imply that setting the above components to zero is 
a consistent truncation of the equations of motion for the action~(\ref{u22gs}). The GS action for the 
truncated theory then is
\ba
{\cal L}_{\mbox{\scriptsize GS } U(1|1)^2/U(1)^2}&=&\frac{1}{2}\int d^2\sigma
\,\,\sqrt{g}g^{\mu\nu}\left( (X^\dg\p_\mu X+Y^\dg\p_\mu Y)(X^\dg\p_\nu X+Y^\dg\p_\nu Y)\right.
\nonumber \\&&\qquad\qquad\,\,\,\,\,\,\,\,\,\,\,\,\,\,\,\,\,\left.
-(\tX^\dg\p_\mu \tX+\tY^\dg\p_\mu \tY)(\tX^\dg\p_\nu \tX+\tY^\dg\p_\nu \tY)
\right)\nonumber \\&&\qquad\,\,\,\,\,\,\,\,\,\,\,\,
-2i\epsilon^{\mu\nu}\left(
X^\dg \p_\mu\tY Y^\dg\p_\nu\tX
+\tX^\dg\p_\mu Y\tY^\dg\p_\nu X\right)\,.\label{gsu112}
\ea
It has two U(1) gauge invariances
\ba
X&\rightarrow& e^{i\theta_1} X\,,\qquad Y\rightarrow e^{i\theta_1} Y\,,\\
\tX&\rightarrow& e^{i\theta_2}\tX\,,\qquad \tY\rightarrow e^{i\theta_2} \tY\,,
\ea
as well as $\kappa$-symmetry which is simply the restriction of equations~(\ref{ku22coord}) and~(\ref{ku22metric}).

\noindent 
If we parametrise the group element $g=(X,Y,\tX,\tY)\in U(1|1)^2$ by
\ba
X&=&(e^{it/2}(1+\frac{1}{2}\psi^2)\,,\,0\,,\,0\,,\,-e^{-i\a/2}\psib)\,,\qquad
Y=(0\,,\,e^{it/2}(1+\frac{1}{2}\eta^2)\,,\,-e^{-i\a/2}\etab\,,\,0)\,,\\
\tX&=&(0\,,\,e^{it/2}\eta\,,\,e^{-i\a/2}(1-\frac{1}{2}\eta^2)\,,\,0)\,,\qquad
\tY=(e^{it/2}\psi\,,\,0\,,\,0\,,\,e^{-i\a/2}(1-\frac{1}{2}\psi^2))\,,
\ea
where $\psi^2\equiv\psib\psi$ and $\eta^2\equiv\etab\eta$, the action~(\ref{gsu112}) becomes
\ba
{\cal L}_{\mbox{\scriptsize GS } U(1|1)^2/U(1)^2}&=&\int d^2\sigma\sqrt{g}g^{\mu\nu}
\left(-\p_\mu\phi_+\p_\nu\phi_-
+i\p_\mu\phi_+\eta_i\overleftrightarrow{\p_\nu}\eta^i
-\p_\mu\phi_+\p_\nu\phi_+\eta^i\eta_i\right)
\nonumber \\&&\qquad\,\,\,\,\,\,\,
-\epsilon^{\mu\nu}\p_\mu\phi_+(\eta_1\overleftrightarrow{\p_\nu}\eta_2-\eta^1\overleftrightarrow{\p_\nu}\eta^2)
\,,\label{gsu112comp}
\ea
This action was postulated in~\cite{aaf} to be a consistent truncation of the full Type IIB GS action on
$AdS_5\times S^5$, by checking the absense of certain cubic terms in the latter action, using an explicit 
non-unitary representation for $PSU(2,2|4)$. Here we have shown that on group-theoretic grounds this action is indeed
such a consistent truncation, and have obtained its form using a unitary representation of the group.

On the local coordinates defined above $\kappa$-symmetry acts as
\be
\delta\eta_i=\epsilon_i\,,\qquad
\delta t=-\delta\a=i\left(\eta^i\epsilon_i+\eta_i\epsilon^i\right)\,.
\ee
In particular notice that $\delta\phi_+=0$. The parameters $\epsilon_i$ are not however free, instead they are
given by
\be
\epsilon_j=\frac{i}{2}\left(\eta_i\overleftrightarrow{\p_\a}\eta^i+i\p_\a\phi_-+i\eta^i\eta_i\p_\a\phi_+\right)
\kappa^\a_j\,.\label{epskappa}
\ee
Above $\kappa^\a_i$ are complex-valued Grassmann functions of the world-sheet; their complex conjugates are
denoted by $\kappa^{\a\,i}$. We will also require that the metric vary under $\kappa$-symmetry as
\ba
\delta(\sqrt{-g} g^{\a\b})&=&-\frac{i}{2}\left[
\kappa^{(\a}P_+^{\b)\gamma}(-\eta_1\p_\gamma\phi_+-i\eta^2\p_\gamma\phi_++2i\p_\gamma\eta_1+2\p_\gamma\eta^2)
\right.\nonumber \\ &&\qquad\!\!\!
+{\bar\kappa^{(\a}}P_+^{\b)\gamma}(-\eta^1\p_\gamma\phi_++i\eta_2\p_\gamma\phi_+-2i\p_\gamma\eta^1
+2\p_\gamma\eta_2)
\nonumber \\ &&\qquad\!\!\!
+{\tilde\kappa^{(\a}}P_-^{\b)\gamma}(\eta_1\p_\gamma\phi_+-i\eta^2\p_\gamma\phi_+-2i\p_\gamma\eta_1
+2\p_\gamma\eta^2)
\nonumber \\ &&\left.\qquad\!\!\!
+{\bar{\tilde\kappa}}^{(\a}P_-^{\b)\gamma}(\eta^1\p_\gamma\phi_++i\eta_2\p_\gamma\phi_++2i\p_\gamma\eta^1
+2\p_\gamma\eta_2)
\right]
\nonumber \\
&=&-\frac{i}{2}\left[
\sqrt{-g}g^{\a\gamma}\left(\kappa^{\b\,i}\p_\gamma\eta_i+\kappa^{\b}_i\p_\gamma\eta^i
+\frac{i}{2}\p_\gamma\phi_+(\kappa^{\b\,i}\eta_i-\kappa^{\b}_i\eta^i)
\right)
\right.\nonumber \\ &&\qquad\!\!\!\left.
+i\epsilon^{\a\gamma}\left(\kappa_1^\b\p_\gamma\eta_2+\kappa_2^\b\p_\gamma\eta_1
-\kappa^{\b\,1}\p_\gamma\eta^2-\kappa^{\b\,2}\p_\gamma\eta^1
\right)\right.\nonumber \\ &&\qquad\!\!\!\left.
-\frac{1}{2}\epsilon^{\a\gamma}\p_\gamma\phi_+
\left(\kappa_1^\b\eta_2+\kappa_2^\b\eta_1
+\kappa^{\b\,1}\eta^2+\kappa^{\b\,2}\eta^1
\right)
\right]\,.
\ea
where $a^{(\a}b^{\b)}=a^\a b^\b+a^\b b^\a$ and
\be
\kappa^\a_1=\frac{i}{2}({\bar{\tilde\kappa}}^\a-{\bar \kappa^\a})\,,\qquad
\kappa^\a_2=\frac{1}{2}({\tilde\kappa}^\a+\kappa^\a)\,,
\ee
with the complex conjugates defined as $\kappa^\dg\equiv {\bar\kappa}$ and ${\tilde\kappa}^\dg\equiv
{\bar{\tilde\kappa}}$.  The above variation of the metric is symmetric
and since $\sqrt{-g} g^{\a\b}$ has unit determinant (is uni-modular) we require that
\be
\kappa^\a=P_+^{\a\b}\kappa_\b\,,
\qquad
{\tilde\kappa}^\a=P_-^{\a\b}{\tilde\kappa}_\b\,.
\ee
Using the above formulas one can check that the action~(\ref{gsu112comp}) is indeed invariant under this symmetry.
However, as we show below this local symmetry is trivial on-shell.

\subsection{Fake $\kappa$-symmetry}\label{partlim}

In this sub-section we show that $\kappa$-symmetry acts trivially on-shell on the fermionic GS actions studied in 
this section. To see this most easily we will first
consider the particle limit (in other words we remove all $\sigma$ dependence of fields) for the action 
${\cal L}_{\mbox{\scriptsize  GS } U(1|1)^2/U(1)^2}$. This gives
\be
{\cal L}_{\mbox{\scriptsize particle}}=-\int d\tau e^{-1}{\dot\phi}_+\left({\dot\phi}_-
+{\dot\phi}_+\eta^i\eta_i-i\eta_i{\dot\eta}^i-i\eta^i{\dot\eta}_i\right)=-\int d\tau e^{-1}{\dot\phi}_+a\,,
\label{supart}
\ee
where for convenience we have defined~\footnote{As an aside note that the fermion index $i$ can now run over
any number and is not restricted to $i=1,2$ as is the case for the super-string.  This is quite typical of
$\kappa$-invariant particle actions. }
\be
a=\left({\dot\phi}_-
+{\dot\phi}_+\eta^i\eta_i-i\eta_i{\dot\eta}^i-i\eta^i{\dot\eta}_i\right)\,.
\ee
Setting $e=\mbox{constant}$, we may solve the the $\phi_+$, $\phi_-$ and $\eta_i$ equations of motion to get
\be
\phi_+=2\kappa\tau\,,\qquad \phi_-=\lambda\tau\,,\qquad \eta_i=e^{-i\kappa\tau}\eta_{0\,i}\,,
\ee
where $\kappa$, $\lambda$ (respectively, $\eta_{0\,i})$ are complex constant Grassmann-even (odd)
numbers.\footnote{In the above solution we have, without loss of generality, set the constant parts of
$\phi_+$ and $\phi_-$ to zero.} Finally, we turn to the equation for the einbein $e$ which reduces to
\be
\kappa\lambda=0\,.
\ee
or in other words forces us to set either $\kappa$ or $\lambda$ to zero. As a result the theory consists of two sectors, 
one with $\kappa=0$ and the other with $\lambda=0$. The former sector is trivial and uninteresting as all fields apart from 
$\phi_-$ are constant and the energy is zero. The physically more relevant sector has $\lambda=0$ and $\kappa\neq 0$.

\noindent Let us now turn to the $\kappa$ invariance of the action~(\ref{supart}). It is easy to see that this
action is invariant under
\ba
\delta\phi_+=0\,,\qquad
\delta\eta_i=a\kappa_i\,,\qquad
\delta\phi_-=ia(\eta^i\kappa_i+\eta_i\kappa^i)\,,\nonumber \\
\delta(e^{-1})=2i({\dot\eta^i}\kappa_i+{\dot\eta_i}\kappa^i)+{\dot\phi}_+(\eta^i\kappa_i+\eta_i\kappa^i)\,,
\ea
where $\kappa_i$ are arbitrary Grassmann-odd functions of $\tau$. Since we are free to pick the parameters
$\kappa_i$ one might think that we could simply gauge away the femrionic degrees of freedom using this symmetry;
had the $\kappa$ variations been of the form
\be
\delta\eta_i=\kappa_i\,,\nonumber
\ee
we would have been able to gauge away the fermions. In fact this is not the case: the $\kappa$ variation of the 
fermions instead reads
\be
\delta\eta_i=a\kappa_i\,,
\ee
From the equation for the einbein $e$ we see that in fact $a=0$ (in the physically important sector for which 
$\kappa\neq  0$ as discussed above) and so on-shell the above $\kappa$ symmetry
acts trivially on all fields except the einbein itself. But any $\kappa$ variation of the einbein
$e$ can be compensated for by a diffeomorphism. We conclude that while the actions~(\ref{supart})
and~(\ref{gsu112}) formally have a $\kappa$-symmetry, this has a trivial action on-shell and so cannot be used
to eliminate any fermions.

\noindent
The argument in the above paragraph relies on the fact that on fermions $\kappa$-symmetry was acting as
$\delta\eta_i=a\kappa_i$ and on-shell $a=0$. Returing to the fermionic GS superstring actions discussed 
in this section we see from equation~(\ref{ku22eps}) that here too $\kappa$-symmetry acts as 
$\delta\eta_i=a_{\mbox{\scriptsize string}}\kappa_i$, where now
\be
a_{\mbox{\scriptsize string}}=
(X^\dg\p_\a X+Y^\dg\p_\a Y+\tX^\dg\p_\a \tX+\tY^\dg\p_\a \tY)\,.
\ee
It is easy to check that because of the Virasoro constrains $a_{\mbox{\scriptsize string}}$ is also zero 
on-shell. We conclude that the $\kappa$-symmetry of the action~(\ref{gsu112comp}) is trivial on-shell and so
cannot be used to eliminate any fermions.

\section{Large charge limits of fermionic GS actions}\label{sec4}
\setcounter{equation}{0}

Given a $\Zop_4$ automorphism on some coset $G/H$ we may construct a Green-Schwarz Lagrangian for it~(\ref{z4gs}).  On
general grounds the large charge limit of this Lagrangian should be a generalised Landau Lifshitz sigma model. Further,
since we expect the global charges of the two actions to map onto one another, this LL sigma model should be
constructed on a coset $G/{\tilde H}$. In this section we will attempt to identify ${\tilde H}$.

One step in this direction is to count the number of degrees of freedom that the GS action has and compare it with that
of the LL model.  For example in the case of the Type IIB superstirng on $AdS_5\times S^5$ there are 10 real bosonic
degrees of freedom, and there are $32/2=16$ fermionic degrees of freedom (where the factor of $1/2$ comes from $\kappa$
symmetry). In the large charge limit two of the bosonic degrees of freedom are eliminated; the remaining eight are
'doubled' since the LL Lagrangian should be thought of as a Lagrangian on phase space. The 16 fermions are described by
coupled first order equations. When taking the LCL we integrate out half of the fermions, in order to arrive at second
order equations~\cite{st2}, leaving us with 8 real fermionic degrees freedom; as in the case of the bosons this should
also be 'doubled', leaving us with 16 fermionic degrees of freedom.  At this point we may simply guess what ${\tilde
H}$ is in the case of $G=PSU(2,2|4)$, since the only coset of the form $G/{\tilde H}$ with 16 bosonic and fermionic 
degrees of freedom each is 
\be
{\tilde H}=PS(U(1,1|2)\times U(2|2))\,, 
\ee 
though of course in this case ${\tilde H}$ is well known from gauge theory. 

Let us persue this counting argument further and consider the GS action on
\be
\frac{U(1|1)^2}{U(1)^2}\,.
\ee
This is a sub-sector of the classical GS string action on $AdS_5\times S^5$. It has 2 real bosonic degrees of freedom
and 4 real fermionic degrees of freedom. As was shown in section~\ref{partlim}, $\kappa$-symmetry in this case is
trivial on-shell, and so, following the counting argument in the previous paragraph,~\footnote{For the bosons we
subtract two real degrees of freedom in the LCL and double the remaining ones.  In the present case this gives
$2\times(2-2)=0$ d.o.f. For the fermions, the number of d.o.f. in the LL sigma model should be the same as that of
the GS string once $\kappa$-symmetry is fixed. This is because, once $\kappa$-symmetry is fixed, we halve the number
of d.o.f. since the GS action gives first order differential equations, and the LL action gives second order
differential equations; we then double it because the LL action is an action on phase space. In the present case,
since $\kappa$-symmetry is trivial on-shell we end up with $2\times 4/2=4$ fermionic d.o.f.} we expect the LL sigma
model corresponding to the LCL of this GS action to have 4 real fermionic degrees of freedom and no bosonic degrees
of freedom. The only such coset is
\be
\frac{U(1|1)^2}{U(1)^4}\,,
\ee
in other words ${\tilde H}=U(1)^4$. 

Similarily, we may consider the bigger sub-sector of the full classical superstring on $AdS_5\times S^5$
\be
\frac{U(2|2)}{SU(2)^2}\,,
\ee
for which $\kappa$-symmetry is also trivial on-shell. This sub-sector has 2 bosonic and 8 fermionic d.o.f. As a result
we expect the LL sigma-model to have no bosonic d.o.f. and 8 fermionic d.o.f. Again this is enough for us to identify
\be
\frac{U(2|2)}{U(2)^2}\,,
\ee
as the coset on which the LL sigma model is constructed. 
Finally, the largest classical sub-sector of the GS
string action on $AdS_5\times S^5$ for which 
$\kappa$-symmetry is trivial is the GS action on
\be
\frac{PS(U(1,1|2)\times U(2|2))}{SU(1,1)\times SU(2)^3}\,.
\ee
By our counting argument the corresponding LCL coset 
should have 16 fermionic and no bosonic d.o.f. As a
result, the LL sigma model which corresponds to the 
LCL limit of the GS action on $(U(1,1|2)\times
U(2|2))/SU(1,1)\times SU(2)^3$ is constructed over the coset
\be
\frac{PS(U(1,1|2)\times U(2|2))}{U(1,1)\times U(2)^3}\,.
\ee

While this counting argument shows how to identify 
${\tilde H}$, it is not very clear how the LCL should be
taken in practice and in particular how starting 
from a GS action one arrives at a LL action. The rest 
of this section will address these issues in the three 
cases of $G=U(1|1)^2\,,\,U(2|2)$ and $U(1,1|2)\times U(2|2)$. 
We will restrict our discusion to the leading order 
term in the LCL and leave the matching of sub-leading 
terms to a future publication.

\subsection{Matching the $U(1|1)^2$ sub-sectors}

In this subsection we will argue that the large charge limit of the Lagrangian given in 
equations~(\ref{gsu112}) and~(\ref{gsu112comp}) which describes the Green-Schwarz string on the coset
\be
\frac{U(1|1)^2}{U(1)^2}\,,
\ee
is given by the Landau-Lifshitz Lagrangian on the coset
~\footnote{This is somewhat different to the 
comparison between gauge and string theory done in~(\cite{aaf}) where it was argued that on the gauge theory
side the coset should be $U(1|1)/U(1)^2$.}
\be
\frac{U(1|1)^2}{U(1)^4}\,.
\ee
We will first arrive at this result in a very pedestrian way. Since general solutions to both the LL and GS
cosets can be given explicitly in full generality we will write them down using unconstrained coordinates. On
the GS side,
\be
\phi_+=\kappa\tau\,,
\ee
the general solution takes the form
\be
\eta_1=\sum_{n=-\infty}^\infty e^{in\sigma}\left(e^{i\omega_n\tau}\psi^+_n +e^{-i\omega_n\tau}\psi^-_n\right)\,,
\ee
where $\psi^\pm_n$ are constant Grassmann-odd numbers, and
\be
\omega_n=\sqrt{n^2+\kappa^2/4}\,.
\ee
$\eta_2$ is completely determined via the equation of motion
\be
\p_\sigma\eta_2=i\p_\tau\eta^1-\frac{\kappa}{2}\eta^1\,.
\ee
In the LCL we take $\kappa\rightarrow\infty$ in which case we have
\ba
\eta_1&\sim&\sum_{n=-\infty}^\infty e^{in\sigma}\left(e^{i(\kappa/2+n^2/\kappa)\tau}\psi^+_n 
+e^{-i(\kappa/2+n^2/\kappa)\tau}\psi^-_n\right)\nonumber \\
&=&e^{i\kappa\tau/2}\left[\psi^+_0+\sum_{n=1}^\infty 
e^{in^2\tau/\kappa}\left(\psi^+_ne^{in\sigma}+\psi^+_{-n}e^{-in\sigma}\right)\right]\nonumber \\
&& +e^{-i\kappa\tau/2}\left[\psi^-_0+\sum_{n=1}^\infty 
e^{-in^2\tau/\kappa}\left(\psi^-_ne^{in\sigma}+\psi^-_{-n}e^{-in\sigma}\right)\right]\nonumber \\
&\equiv& e^{i\kappa\tau/2}\psi_{1\,\,\mbox{\scriptsize 
LL}}+e^{-i\kappa\tau/2}\psib_{2\,\,\mbox{\scriptsize LL}}\,,
\ea
where $\psi_{1\,\,\mbox{\scriptsize LL}}$ and $\psi_{2\,\,\mbox{\scriptsize LL}}$ are the 2 
complex fermionic d.o.f. for the LL sigma model on (see equation~(\ref{LLu11}))
\be
\frac{U(1|1)^2}{U(1)^4}\,.
\ee
In particular, after rescaling $\tau\rightarrow\kappa\tau$, they satisfy the 
equations of motion
\be
0=\left(\p_\sigma^2- i\p_\tau\right)\psi_{1,2\,\,\mbox{\scriptsize LL}}\,.
\ee
In this way we match, to leading order in the LCL, the classical string Lagrangian with the 
corresponding 
coherent state continuum limit of the gauge theory dilatation operator in the $U(1|1)^2$ 
sub-sector. 

Notice that physical string solutions have to satisfy the level-matching condition
\be
\int_0^{2\pi}\p_1\phi_-=2\pi m\,,\qquad\mbox{for }m\in \Zop\,.\label{levmatch}
\ee
The winding parameter $m$ does not, however, enter the LCL Lagrangian
Rather, it gives a constraint on its solutions. This matches the spin-chain side where
$m$ enters as a constraint on the Bethe roots, but does not enter the algebraic Bethe
equations or the LL sigma-model action. This feature is very similar to the
$SL(2)$ sector discussed in~\cite{ptt}.

\subsection{Large Charge Limit of fermionic GS actions}

In this section we re-phrase the above discussion in terms of the 
embedding coordinates $X,\, Y\dots$, and the currents $j^{(k)}_\mu$. 
This allows for a straightforward generalisation from the $U(1|1)^2$ 
sub-sector to the $U(2|2)$ and $U(1,1|2)\times U(2|2)$ sub-sectors. 
We present the explicit discussion only for the case of $U(2|2)$, 
but the other case follows almost trivially.

The first thing to note is that 
%% all three of the GS cosets have only 
%% two bosonic coordinates. If fermions are set to zero, the geodesics 
%% are described by $U(1)\times U(1)$ motions. Hence, we are motivated to define
%% \be
%% X_i=e^{it(\tau,\sigma)}A_i\,,\qquad
%% Y_i=e^{it(\tau,\sigma)}B_i\,,\qquad
%% \tX_i=e^{i\a(\tau,\sigma)}\tA_i\,,\qquad
%% \tY_i=e^{i\a(\tau,\sigma)}\tB_i\,,
%% \ee
%% where $A,\,B,\,\tA,\,\tB$ satisfy the same orthonormality 
%% relations as $X,\,Y,\,\tX,\,\tY$ given in
%% equation~(\ref{ads2s2const}). In the Large Charge Limit we then consider
%% \be
%% t=\kappa\tau\,,\qquad\a=-\kappa\tau+\dots\,,
%% \ee
%% In fact,}
the equation of motion for one of the two bosonic fields, $\phi_+$, 
is particularily simple in the GS 
models presently considered. This can be obtained as the super-trace of equation~(\ref{eom1}). As a result we 
may set
\be
X^\dg\p_\mu X+Y^\dg\p_\mu Y-\tX^\dg\p_\mu\tX-\tY^\dg\p_\mu\tY=i\kappa\delta_{\mu\,,\,,0}\,.
\label{phipansatz}
\ee
Using this, in conformal gauge the equation of motion for the off-diagonal component of the
worldsheet metric implies that
\be
X^\dg\p_\sigma X+Y^\dg\p_\sigma Y+\tX^\dg\p_\sigma\tX+\tY^\dg\p_\sigma\tY=0\,,\label{offdiagVir}
\ee
while the fermionic equations of motion~(\ref{eom2}),~(\ref{eom3}) reduce to~\footnote{In terms of 
$X,\,Y,\,\tX,\,\tY$ this implies that we have relations of the form
\be
X^\dg\p_\tau\tY=i\tX^\dg\p_\sigma Y\,,\qquad
\tX^\dg\p_\tau Y=-iX^\dg\p_\sigma\tY\,,\qquad \mbox{{\it etc}}\,.\nonumber 
\ee}
\be
0=\kappa(j^{(3)}_\tau-j^{(3)}_\sigma)+\dots\,,\qquad
0=\kappa(j^{(1)}_\tau+j^{(1)}_\sigma)+\dots\,.\label{fermrel}
\ee
As a result of these relations the WZ term does not 
contribute to the bosonic equation of motion~(\ref{eom1}).
\footnote{This is easy to see since 
the WZ term's contribution to these equations is proportional to 
$\left[ j^{(1)}_\tau\,,j^{(1)}_\sigma\right]-\left[ j^{(3)}_\tau\,,j^{(3)}_\sigma\right]$. 
However, since $j^{(1)}_\tau=-j^{(1)}_\sigma$ and $j^{(3)}_\tau=j^{(3)}_\sigma$ each of these
commutators vanishes seperately.}
This fact allows us to check explicitly that the bosonic 
equations of motion, together with the ansatz~(\ref{phipansatz}), are consistent with 
the equations of motion for the metric $g_{\mu\nu}$ in conformal gauge. In fact these Virasoro 
constraints then imply that
\be
D_\mu t =\delta_{\mu\,,0}\frac{\kappa}{2}\,,\qquad
{\tilde D}_\mu\a =-\delta_{\mu\,,0}\frac{\kappa}{2}\,.
\ee
As in the discussion around equation~(\ref{levmatch}) above, 
the level matching condition that follows from the Virasoro constraints
does not enter the LCL action.

Using equations~(\ref{phipansatz}),~(\ref{offdiagVir}) and~(\ref{fermrel}) together
with a rescaling $\tau\rightarrow\kappa\tau$ we may re-write the GS Lagrangian in 
conformal gauge as follows
\ba
{\cal L_{\mbox{\scriptsize GS $U(2|2)/SU(2)^2$}}}&=&\eta^{\mu\nu}\Str(j^{(2)}_\mu j^{(2)}_\nu)+                                                                                                     
\epsilon^{\mu\nu}\Str(j^{(1)}_\mu j^{(3)}_\nu)\nonumber \\
&=&\eta^{\mu\nu}
\left(X^\dg\p_\mu X+Y^\dg\p_\mu Y-\tX^\dg\p_\mu\tX-\tY^\dg\p_\mu\tY\right)
\nonumber \\&&\qquad\times\,\,
\left(X^\dg\p_\mu X+Y^\dg\p_\mu Y+\tX^\dg\p_\mu\tX+\tY^\dg\p_\mu\tY\right)
\nonumber \\ &&
-2\Str\left(j^{(1)}_\sigma j^{(3)}_\sigma\right)
\nonumber \\
&=&
i\left(X^\dg\p_\tau X+Y^\dg\p_\tau Y+\tX^\dg\p_\tau\tX+\tY^\dg\p_\tau\tY
\right)
-\str\left((j^{(1)}_\sigma +j^{(3)}_\sigma)(j^{(1)}_\sigma +j^{(3)}_\sigma)
\right)\nonumber \\
&=&{\cal L_{\mbox{\scriptsize LL $U(2|2)/U(2)^2$}}}
\ea
The right-hand side of the above equation is nothing but the LL sigma model
Lagrangian defined on $G/{\tilde H}$, where
${\tilde H}$ is fixed under the $\Zop_2$ automorphism which is the square of the $\Zop_4$
automorphism used in the construction of the GS action. We have thus shown that to 
leading order in the LCL the fermionic GS actions constructed in section~\ref{sec3} above reduce to LL
sigma model actions in the manner anticipated by the general argument presented at the start of the present section.
It would be interesting to consider sub-leading corrections to this LCL for example in a manner similar 
to~\cite{k3}.

\subsection{A gauge-theory inspired $\kappa$ gauge}

The GS sigma model on $AdS_5\times S^5$ has $\kappa$-symmetry. This, as well as other symmetries of the string action, 
such as world-sheet diffeomorphisms, are not manifest in the corresponding spin-chain simply because this latter 
system keeps track only of the physical degrees of freedom. One of the challenges of defining a LCL is to identify 
suitable gauges for these stringy symmetries in which the physical degrees of freedom are written in the most natural 
coordinates for the spin-chain: while all gauges should be in principle equivalent it may be much more difficult to 
define a LCL between the two theories if we pick an unnatural gauge. In the previous sub-section we have defined an 
LCL which matches all 16 fermionic degrees of freedom from the GS action to the corresponding LL model in a 
very natural way. This strongly suggests what $\kappa$-gauge should be used in the full $AdS_5\times S^5$ string 
action when comparing to gauge theory. Specifically it should be the gauge which keeps non-zero the 16 fermions of the 
coset $PS(U(1,1|2)\times U(2|2))/(SU(1,1)\times SU(2)^3)$. In fact this is the gauge used recently in~\cite{fpz} and 
the above argument can be interpreted as one motivation for their $\kappa$-gauge choice.

\section*{Acknowledgements}

I am grateful to Arkady Tseytlin for many stimulating discussions throughout this project and to Chris Hull
for a number of detailed conversations on $\kappa$-symmetry. I would also like to thanks Charles Young 
for 
discussions and Gleb Arutyunov for providing a copy of 
his notes~\cite{glebnotes}. This research is funded by EPSRC and MCOIF.

\appendix

\section{Some examples of Landau-Lifshitz sigma models}\label{appe}
\setcounter{equation}{0}
                                                                                                          
In this appendix we collect some expressions for a number of relevant
Landau-Lifshitz sigma models.
                                                                                                          
\subsection{The $SU(2|3)/S(U(2|2)\times U(1))$ model}
                                                                                                          
\noindent
The $SU(2|3)$ sub-sector sigma model Lagrangian is~\cite{st1}
\be
{\cal L}_{\mbox{\scriptsize LL SU(2$|$3)}}=-iU^i\p_\tau U_i-i\psi^\a\p_\tau\psi_a
-\frac{1}{2}|D_\sigma U_i|^2-\frac{1}{2}\Db_\sigma\psi^aD_\sigma\psi_a
+\Lambda(U_iU^i+\psi_a\psi^a-1)\,,\label{gtsu23}
\ee
where
\be
D_\mu\equiv\p_\mu -iC_\mu \,,\qquad \Db_\mu\equiv\p_\mu +iC_\mu \,,\qquad
C_\mu=-iU^i\p_\mu U_i-i\psi^a\p_\mu\psi_a\,,
\ee
and $\psi^a=\psi_a^*$ and $a=1,2$.
                                                                                                          
\subsection{The $SU(4)/S(U(2)\times U(2))$ model}\label{so6}
                                                                                                          
\noindent
The $SO(6)\sim SU(4)$ sub-sector sigma model Lagrangian is~\cite{st1}
\ba
{\cal L}_{\mbox{\scriptsize LL SU(4)}}
&=&{\cal L}_{\mbox{\scriptsize SU(4) WZ}}-\frac{1}{8}\Tr(\p_1 m)^2
-\frac{1}{32}\Tr(m\p_1m)^2+\Lambda(m-m^3)
\nonumber \\
&=&-iV^i\p_\tau V_i-\frac{1}{2}|D_\sigma V_i|^2
+\Lambda_1(V^iV_i-1)+\Lambda_2(V_iV_i-1)+\Lambda_2^*(V^iV^i-1)\,,\label{gtso6}
\ea
where $m_{ij}$ is a $6\times 6$ matrix, related to $V_i$ by
\be
m_{ij}=V_iV^j-V_jV^i\,,
\ee
and
\be
D_\mu \equiv\p_\mu -iC_\mu \,,\qquad C_\mu=-iV^i\p_\mu V_i\,.
\ee
Let us define
\be
M^{A}{}_B=\frac{1}{2}m_{ij}\rho^{ijA}{}_B\,,\qquad
m_{ij}=\frac{1}{4}\tr(M\rho^{ij})\,,
\ee
where $\rho$ are the usual $SU(4)$ $\rho$-matrices.
Notice that
\be
\Tr M= 0\,,\qquad M^\dg=M\,,\qquad M^2=M\,.
\ee
and so we can write it as
\be
M=2XX^\dg-1=-2YY^\dg+1\,,
\ee
where now $X$ and $Y$ are $4\times 2$ matrices which satisfy
\ba
X^\dagger X ={\bf 1}_2\,,\qquad Y^\dagger Y={\bf 1}_2\,,&\qquad&
X^\dagger Y=0\,,\qquad Y^\dagger X=0\,,\\
XX^\dagger+YY^\dagger &\!\!\!\!\!\!\!\!=&\!\!\!\!\!\!\!\!{\bf 1}_4\,.
\ea
Further we can write the $4\times 2$ matrix $X$ as two four-component vectors $u_A$ and $v_A$
\be
X=(u_A,v_A)\,,\label{collvect}
\ee
in terms of which $M^A{}_B$ can be written as
\be
M^A{}_B=2u^Au_B+2v^Av_B-\delta^A_B\,,
\ee
with
\be
u^Au_A=1\,,\qquad v^Av_A=1\,,\qquad u^Av_A=0\,.\label{uvconds}
\ee
We can relate $u_A$ and $v_A$ to $V_i$ by
\be
V_i=\frac{1}{\sqrt{2}}u^A\rho^{i}_{AB}v^B\,,\qquad
V^i=\frac{1}{\sqrt{2}}v_A\rho^{iAB}u_B\,.\label{natgencoords}
\ee
It is an easy check to see that these are consistent with
\be
V_iV^i=1\,,\qquad V_iV_i=0\,,\qquad M^A{}_B=V_iV^j\rho^{ijA}{}_B\,.
\ee
In terms of these, the Lagrangian is
\ba
{\cal L}_{\mbox{\scriptsize LL SU(4)}}
&=&-iu^A\p_0u_A-iv^A\p_0v_A
-\frac{1}{2}\Bigl(\p_1u^A\p_1u_A+\p_1v^A\p_1v_A
\nonumber \\&&
+u^A\p_1u_Au^B\p_1u_B+v^A\p_1v_Av^B\p_1v_B
+2u^A\p_1v_Av^B\p_1u_B\Bigr)
\nonumber \\
&=&-i\Tr(X^\dg\p_0X)-\frac{1}{2}\Tr(\Db_1X^\dg D_1X)\,,\label{goodllso6}
\ea
As before
\ba
X&=&(u_A,v_A)\,,\qquad X^\dg\equiv \left(\begin{array}{c}u^A\\v^A\end{array}\right)\,,
\\
D_\mu X&=&\p_\mu X-X X^\dg\p_\mu X\,.
\ea
The action~(\ref{goodllso6}) has a local U(2) invariance
\be
X\rightarrow XU(\tau,\sigma)\,,
\ee
for $U(\tau,\sigma)$ a general U(2) matrix
\be
U^\dg(\tau,\sigma)U(\tau,\sigma)=U(\tau,\sigma)U^\dagger(\tau,\sigma)=1_2\,.
\ee
In terms of the $u_A$ and $v_A$ the action~(\ref{goodllso6}) is invariant with respect to the
following local transformations
\ba
(u_A,v_A)&\rightarrow&(\cos\theta(\tau,\sigma)\, u_A+\sin\theta(\tau,\sigma)\, v_A,
-\sin\theta(\tau,\sigma)\, u_A +\cos\theta(\tau,\sigma)\, v_A)\,,\nonumber \\
(u_A,v_A)&\rightarrow&(e^{i\phi_1(\tau,\sigma)}u_A,e^{i\phi_1(\tau,\sigma)}v_A)\,,\nonumber \\
(u_A,v_A)&\rightarrow&(e^{i\phi_2(\tau,\sigma)}u_A,e^{-i\phi_2(\tau,\sigma)}v_A)\,,\nonumber \\
(u_A,v_A)&\rightarrow&(e^{i\phi_3(\tau,\sigma)}v_A,-e^{i\phi_3(\tau,\sigma)}u_A)\,,
\ea
                                                                                                          
\subsubsection{Subsectors of the $SU(4)/S(U(2)\times U(2))$ model}\label{so6sub}
                                                                                                          
When written in terms of the $V_i$, the Lagrangian
${\cal L}_{\mbox{\scriptsize SU(4)}}$ can be reduced to the SU(3)
sub-sector by requiring
\be
V^{2a}=-iV^{2a-1}\equiv \frac{1}{\sqrt{2}}U^a\,,\qquad a=1,2,3\,,
\ee
which can further be restriced to the SU(2) subsector for $V^5=0=V^6$. In
terms of the $u_A$ and $v_A$ this restriction is easily enforced by setting
for example
\be
u_A=(U_1,U_2,U_3,0)\,,\qquad
v_A=(0,0,0,1)\,.
\ee
Since $U_aU^a=1$, this choice satisfies the constraints~(\ref{uvconds}).
Restricting to the SU(2) sector is achieved by setting $u_3=U_3=0$.
Upon inserting these ansatze, the Lagrangian~(\ref{goodllso6})
reduces to the Lagrangian~(\ref{gtsu3}).
                                                                                                          
\noindent
Another interesting sub-sector is obtained by setting
\be
u_A=(U_1,U_2,0,0)\,,\qquad
v_A=(0,0,V_3,V_4)\,,\label{su2su2sub}
\ee
together with the conditions
\be
U^1U_1+U^2U_2=1\,,\qquad
V^1V_1+V^2V_2=1\,.
\ee
This results in SU(2)$\times$SU(2) subsector consisting of two decoupled SU(2)
Landau Lifshitz Lagrangians.

\subsection{The $SU(2,2)/S(U(2)\times U(2))$ model}\label{so24}
                                                                                                          
For later convenience we present here the
$SO(2,4)/S(O(2)\times O(4))\sim SU(2,2)/S(U(2)\times U(2))$ Landau-Lifshitz
Lagrangian
\ba
{\cal L}_{\mbox{\scriptsize LL SU(2,2)}}&=&-i\tV^i\p_0\tV_i-\frac{1}{2}|D_\sigma \tV_i|^2
\nonumber \\
&=&-i\tu^A\p_0\tu_A-i\tv^A\p_0\tv_A
-\frac{1}{2}\Bigl(\p_1\tu^A\p_1\tu_A+\p_1\tv^A\p_1\tv_A
\nonumber \\&&
-\tu^A\p_1\tu_A\tu^B\p_1\tu_B-\tv^A\p_1\tv_A\tv^B\p_1tv_B
-2\tu^A\p_1\tv_A\tv^B\p_1\tu_B\Bigr)
\nonumber \\
&=&i\Tr(\tX^\dg\p_0\tX)+\frac{1}{2}\Tr(\Db_1\tX^\dg D_1\tX)\,,\label{goodllso24}
\ea
where
\be
\tV^i\equiv \tV_j^*\eta^{ji}\,,\qquad
\mbox{where}\qquad
 \eta_{ij}=\diag(-1,-1,1,1,1,1)\,,
\ee
and
\be
\tu^A\equiv \tu_B^*C^{BA}\,,\qquad
\tv^A\equiv \tv_B^*C^{BA}\,,\qquad
\mbox{where}\qquad
C_{AB}=(1,1,-1,-1)\,.
\ee
The $4\times 2$ matrix $\tX$ has two columns
\be
\tX=(\tu_A,\tv_A)\,,
\ee
and the covariant derivatives are
\ba
D_\mu \tV_i&=&\p_\mu \tV_i +\tV^j\p_\mu \tV_j \tV_i\,, \\
D_\mu \tX&=&\p_\mu \tX-\tX \tX^\dg\p_\mu \tX\,.
\ea
We define
\be
\tX^\dg\equiv-\left(\begin{array}{c}\tu^A\\\tv^A\end{array}\right)\,.
\ee
This is done for convenience, so that the form of the action in terms of $X$ is independent of
the signature. The fields in the Lagrangian~(\ref{goodllso24}) now satisfy the constraints
\ba
\tX^\dg \tX&=&1_2\,,\\
\tV^i\tV_i&=&-1\,,\qquad \tV_i\tV_i=0\,,\\
\tu^A\tu_A&=&-1\,,\qquad \tv^A\tv_A=-1\,,\qquad \tu^A\tv_A=0\,.\label{uvlorconst}
\ea
The action~(\ref{goodllso24}) has a local non-compact U(2) invariance
\be
\tX\rightarrow \tX U(\tau,\sigma)\,,
\ee
for $U(\tau,\sigma)$ a general U(2) matrix
\be
U^\dg(\tau,\sigma)U(\tau,\sigma)=U(\tau,\sigma)U^\dagger(\tau,\sigma)=1_2\,.
\ee
In terms of the $\tu_A$ and $\tv_A$ the action~(\ref{goodllso24}) is invariant with respect to the
following local transformations
\ba
(\tu_A,\tv_A)&\rightarrow&(\cos\theta(\tau,\sigma)\, \tu_A+\sin\theta(\tau,\sigma)\, \tv_A,
-\sin\theta(\tau,\sigma)\, \tu_A +\cos\theta(\tau,\sigma)\, \tv_A)\,,\nonumber \\
(\tu_A,\tv_A)&\rightarrow&(e^{i\phi_1(\tau,\sigma)}\tu_A,e^{i\phi_1(\tau,\sigma)}\tv_A)\,,\nonumber \\
(\tu_A,\tv_A)&\rightarrow&(e^{i\phi_2(\tau,\sigma)}\tu_A,e^{-i\phi_2(\tau,\sigma)}\tv_A)\,,\nonumber \\
(\tu_A,\tv_A)&\rightarrow&(e^{i\phi_3(\tau,\sigma)}\tv_A,-e^{i\phi_3(\tau,\sigma)}\tu_A)\,,
\ea
                                                                                                          
\noindent
To relate the $\tV_i$ coordinates to the $\tu_A$, $\tv_A$ coordinates
recall that the SU(4) $\rho$ matrices could be combined into
$8\times 8$ $\gamma$ matrices of $SO(6)$ as follows
\be
\gamma^i=\left(\begin{array}{cc}0&\rho^i_{AB}\\\rho^{iAB}&0\end{array}\right)\,,
\qquad i=1,\dots,6\,,
\ee
with the $\gamma^i$ satisfying the $SO(6)$ anti-commutation relations
\be
\left\{\gamma^i,\gamma^j\right\}=2\delta^{ij}\,.\label{dirso6}
\ee
The $SO(2,4)$ $\gamma$-matrix algebra is instead
\be
\left\{\tgamma^i,\tgamma^j\right\}=-2\eta^{ij}\,.\label{dirso24}
\ee
Given a set of SO(6) $\gamma$ matrices we can define
\be
\tgamma^i=\left\{\begin{array}{ll} \gamma^i\,,
\qquad &
i=1,2\,,\\
i\gamma^i\,,
\qquad &
i=3,\dots,6\,,\end{array}\right.
\ee
which satisfy~(\ref{dirso24}). Similarily we will define
\be
\trho^i_{AB}=\left\{\begin{array}{ll} \rho^i_{AB}\,,
\qquad &
i=1,2\,,\\
i\rho^i_{AB}\,,
\qquad &
i=3,\dots,6\,,\end{array}\right.\qquad\mbox{and}\qquad
\trho^{iAB}=\left\{\begin{array}{ll} \rho^{iAB}\,,
\qquad &
i=1,2\,,\\
i\rho^{iAB}\,,
\qquad &
i=3,\dots,6\,,\end{array}\right.
\ee
which now satisfy
\be
\trho^i_{AB}\trho^{jBC}+\trho^i_{AB}\trho^{jBC}=-2\delta_A^C\eta^{ij}\,,
\ee
as well as
\be
\eta_{ij}\trho^{i}_{AB}\trho^{jCD}=2(\delta_A^C\delta_B^D-\delta_A^D\delta_B^C)\,.
\ee
Note also that for SU(4) $\rho$ matrices we had
\be
(\rho^i_{AB})^*=-\rho^{iAB}\,,
\ee
while for the SU(2,2) $\trho$ matrices we have
\be
(\trho^i_{AB})^*=\eta^{ij}\rho^{jAB}\,.
\ee
The relationship between the $\tv_A$, $\tu_A$ and the $\tV_i$ is
\be
\tV_i=\frac{1}{\sqrt{2}}\tu^A\trho^{i}_{AB}\tv^B\,,\qquad
\tV^i=\frac{1}{\sqrt{2}}\tv_A\trho^{iAB}\tu_B\,.
\ee
This can be used to derive the equality between the first and second lines in
equation~(\ref{goodllso24}).
                                                                                                          
\subsubsection{Subsectors of the $SU(2,2)/S(U(2)\times U(2))$ model}\label{so24sub}
                                                                                                          
When written in terms of the $V_i$, the Lagrangian
${\cal L}_{\mbox{\scriptsize LL SU(2,2)}}$ can be reduced to the SU(1,2)
sub-sector by requiring
\be
\tV^{2a}=-i\tV^{2a-1}\equiv \frac{1}{\sqrt{2}}\tU^a\,,\qquad a=1,2,3\,,
\ee
which can further be restriced to the SU(2) subsector for $\tV^5=0=\tV^6$. In
terms of the $\tu_A$ and $\tv_A$ this restriction is easily enforced by setting
for example
\be
\tu_A=(\tU_1,\tU_2,\tU_3,0)\,,\qquad
\tv_A=(0,0,0,1)\,.
\ee
We require
\be
\sum_{a=1}^3\eta^{ab}\tU^*_a\tU_b=-1\,,
\ee
so as to satisfy the constraints~(\ref{uvlorconst}).
Restricting to the SU(1,1) sector is achieved by setting $\tu_3=\tU_3=0$.
Upon inserting these ansatze, the Lagrangian~(\ref{goodllso24})
reduces to the standard SU(1,2) Landau-Lifshitz Lagrangian~\cite{st1}
\be
{\cal L}_{\mbox{\scriptsize SU(1,2)}}=-i\tU^a\p_0\tU_a-\frac{1}{2}|D_\sigma \tU_a|^2
+\Lambda(\tU^a\tU_a+1)\,,
\ee
with $a=1,2,3$ and $\tU^a\equiv \eta^{ab}\tU^*_b$.

\subsection{The $SU(2|2)/S(U(1|1)\times U(1|1))$ model}
                                                                                                          
Lets construct the LL model on $SU(2|2)/S(U(1|1)\times U(1|1))$. Starting from
equation~(\ref{LLsigmamodel}), with $\Tr$ now replaced by $\STr$ we may define
\be
g=(X,Y)\,,
\ee
with $X$ and $Y$ super-matrices which satisfy
\be
X^\dg X=1_2\,,\qquad
Y^\dg Y=1_2\,,\qquad
XX^\dg+ YY^\dg=\left(\begin{array}{cc}1_2&0\\0&1_2\end{array}\right)\,.
\ee
The LL Lagrangian for this model is then
\ba
{\cal L}_{\mbox{\scriptsize LL SU(2$|$2)}}
&=&\frac{i}{2}\STr\left[\left(\begin{array}{cc}1_n&0\\0&1_m\end{array}\right)g^{-1}\p_0 g\right]
-\frac{1}{4}\STr\left((g^{-1}D_1g)(g^{-1}D_1g)\right)
\nonumber \\
&=&
i\STr(X^\dg\p_0X)
-\frac{1}{2}\STr\left[\Db_1X^\dg D_1X\right]\,,
\ea
where
\be
D_1X\equiv \p_1 X-XX^\dg\p_1X\,.
\ee
The bosonic base of SU(2$|$2) is SU(2)$\times$SU(2), where in the case of interest to us we write
\be
X=(\tu_A,v_A)\,,\qquad A=1\,\dots,4\,,
\ee
with
\be
\tu^A\tu_A=-1\,,\qquad
v^Av_A=1\,,\qquad
\ee
and
\be
\tu^A\equiv \tu_B^*C^{BA}\,,\qquad
v^A\equiv v_B^*C^{BA}\,,\qquad
X^\dg\equiv-\left(\begin{array}{c}\tu^A\\v^A\end{array}\right)\,,
\ee
where $C^{BA}\equiv \diag(-1,-1,1,1)$.
Note that the first (last) two components of $u_A$ ($v_A$) are bosonic
and the last (first) two components of $u_A$ ($v_A$) are fermionic.
\ba
{\cal L}_{\mbox{\scriptsize LL SU(2$|$2)}}
&=&
-i\tu^A\p_0\tu_A-iv^A\p_0v_A
-\frac{1}{2}\Bigl(\p_1\tu^A\p_1\tu_A+\p_1v^A\p_1v_A
\nonumber \\&&
-\tu^A\p_1\tu_A\tu^B\p_1\tu_B+v^A\p_1v_Av^B\p_1v_B
+2\tu^A\p_1v_Av^B\p_1\tu_B\Bigr)
\,.\label{goodllsu22}
\ea
The action~(\ref{goodllsu22}) has a local non-compact U(1$|$1) invariance
\be
\tX\rightarrow \tX U(\tau,\sigma)\,,
\ee
for $U(\tau,\sigma)$ a general U(1$|$1) matrix
\be
U^\dg(\tau,\sigma)U(\tau,\sigma)=U(\tau,\sigma)U^\dagger(\tau,\sigma)=1_2\,.
\ee
In terms of the $\tu_A$ and $v_A$ the action~(\ref{goodllsu22}) is invariant with respect to the
following local transformations
\ba
(\tu_A,v_A)&\rightarrow&(\tu_A+ v_A\theta_1(\tau,\sigma),
v_A+\tu_A\theta_1(\tau,\sigma))\,,\nonumber \\
(\tu_A,v_A)&\rightarrow&(\tu_A-i v_A\theta_2(\tau,\sigma),
v_A+i\tu_A\theta_2(\tau,\sigma))\,,\nonumber \\
(\tu_A,v_A)&\rightarrow&(e^{i\phi_1(\tau,\sigma)}\tu_A,e^{i\phi_1(\tau,\sigma)}v_A)\,,\nonumber \\
(\tu_A,v_A)&\rightarrow&(e^{i\phi_2(\tau,\sigma)}\tu_A,e^{-i\phi_2(\tau,\sigma)}v_A)\,,
\ea
where $\psi_1$, $\psi_2$ ($\theta_1$, $\theta_2$) are real Grassmann-even (-odd) valued function.

\section{Quantising the action~(\ref{gsu112}) in the $t+\a=\kappa\tau$ gauge}\label{appa}
\setcounter{equation}{0}

Given the simple form of the action~(\ref{gsu112}),~(\ref{gsu112comp}) we present a brief light-cone quantisation of
it here. The main point is that, as expected, the Hamiltonian has a non-zero normal ordering
constant~(\ref{normordconst}).

%Starting from the action~(\ref{aaf}) we may redefine the fermions as in~\cite{aaf}
%\ba
%\eta&=&e^{i(\a+t)} \eta_1\,,\qquad \psi =e^{i(\a+t)} \eta_2\,,\nonumber \\
%\etab&=&e^{-i(\a+t)} \eta^1\,,\qquad \psib =e^{-i(\a+t)} \eta^2\,,
%\ea
%as well as define
%\be
%\phi_\pm=t\pm\a\,.
%\ee
%The action then becomes
%\ba
%{\cal L}_\kappa&=&\int d^2\sigma\sqrt{g}g^{\mu\nu}
%\left(-4\p_\mu\phi_+\p_\nu\phi_-
%+2i\p_\mu\phi_+\eta_i\overleftrightarrow{\p_\nu}\eta^i
%-4\p_\mu\phi_+\p_\nu\phi_+\eta^i\eta_i\right)
%\nonumber \\&&\qquad\,\,\,\,\,\,\,
%-2\epsilon^{\mu\nu}\p_\mu\phi_+(\eta_1\overleftrightarrow{\p_\nu}\eta_2-\eta^1\overleftrightarrow{\p_\nu}\eta^2)
%\,,
%\ea
%where $i=1,2$ and $\lambda\overleftrightarrow{\p}\xi\equiv\lambda\p\xi+\xi\p\lambda$. We may also redefine 
%$\phi_-\rightarrow -\phi_-$ and $\phi_+\rightarrow -\phi_+/2$ to arrive at
%\ba
%-2{\cal L}_\kappa&=&\int d^2\sigma\sqrt{g}g^{\mu\nu}
%\left(\p_\mu\phi_+\p_\nu\phi_-
%+\frac{i}{2}\p_\mu\phi_+\eta_i\overleftrightarrow{\p_\nu}\eta^i
%+\frac{1}{2}\p_\mu\phi_+\p_\nu\phi_+\eta^i\eta_i\right)
%\nonumber \\&&\qquad\,\,\,\,\,\,\,
%-\frac{1}{2}\epsilon^{\mu\nu}\p_\mu\phi_+(\eta_1\overleftrightarrow{\p_\nu}\eta_2
%-\eta^1\overleftrightarrow{\p_\nu}\eta^2)
%\,.
%\ea
Since the equation of motion for $\phi_+$ is
\be
0=\p_\mu\left(\sqrt{g}g^{\mu\nu}\p_\nu\phi_+\right)\,,
\ee
we may impose conformal gauge ($g_{\mu\nu}=\eta_{\mu\nu}$) and set
\be
\phi_+=2\kappa \tau\,.
\ee
The fermionic equations of motion then reduce to 
\ba
0&=&(i\p_0+\kappa)\eta^i+\p_1\eta_j\,,\label{fermeom1}
\ea
where $i\neq j$. The fermionic fields have the
following periodicity conditions
\be
\eta_1(\tau\,,2\pi)=e^{i\alpha}\eta_1(\tau\,,0)\,,\qquad
\eta_2(\tau\,,2\pi)=e^{-i\alpha}\eta_1(\tau\,,0)\,,
\ee
The fermionic equations of motion then are solved by
\ba
\eta_1&=&\sum_{n=-\infty}^\infty\theta_n e^{i(n\sigma+\omega_n\tau)}
+\thetat_n e^{i(n\sigma-\omega_n\tau)}\,,\\
\eta_2&=&\sum_{n=-\infty}^\infty \xi_n e^{-i(n\sigma+\omega_n\tau)}
+\xit_n e^{-i(n\sigma-\omega_n\tau)}\,,
\ea
where 
\be
\omega_n=\sqrt{n^2+\frac{\kappa^2}{4}}\,,
\ee
and for $n\neq 0$
\be
\thetab=\frac{in}{\omega_n+\kappa}\xi_n\,,\qquad
\thetatb=\frac{-in}{\omega_n-\kappa}\xit_n\,,
\ee
while for $n=0$
\be
0=\theta_0=\xit_0\,.
\ee
The Virasoro constraints can be used to find $\phi_-$ in terms of the other fields
\ba
0&=&\p_0\phi_-+\frac{i}{2}\eta_i\overleftrightarrow{\p_0}\eta^i-\frac{\kappa}{2}\eta^i\eta_i\,,\\
0&=&\p_1\phi_-+\frac{i}{2}\eta_i\overleftrightarrow{\p_1}\eta^i\,.
\ea
The N\"other current for time translations $t\rightarrow t+\epsilon$ is
\be
j^t_\mu=-\p_\mu\phi_+-\p_\mu\phi_-+i\eta_i\overleftrightarrow{\p_\mu}\eta^i-2\p_\mu\phi_+\eta^i\eta_i
-\epsilon_\mu{}^\nu\left(\eta_1\overleftrightarrow{\p_\nu}\eta_2
-\eta^1\overleftrightarrow{\p_\nu}\eta^2\right)\,.
\ee
We can use the equations of motion to write the Hamiltonian of the system as
\be
H_\kappa\equiv-\frac{1}{2\pi}\int d\sigma j^t_0
=2\kappa+\frac{i}{2\pi}\int d\sigma \eta_i\overleftrightarrow{\p_0}\eta^i\,.
\label{Hkappa}
\ee

\noindent
The canonical momentum conjugate to $\eta_i$ is $4i\kappa\eta^i$ and so upon quantisation we must have
\be
\left\{\eta^i(\tau,\sigma),\eta_j(\tau,\sigma^\prime)\right\}=-\frac{1}{4\kappa}\delta^i_j\delta(\sigma-\sigma^\prime)\,.
\ee
As a consequence the mode oscillators have the following non-zero anti-commutators
\be
\left\{\xib_n,\xi_m\right\}=-\delta_{nm}\frac{\omega_n+\kappa}{16\pi\kappa\omega_n}\,,\qquad
\left\{\xitb_n,\xit_m\right\}=-\delta_{nm}\frac{\omega_n-\kappa}{16\pi\kappa\omega_n}\,,
\ee
together with
\be
\left\{\xib_0,\xi_0\right\}=-\frac{1}{8\pi\kappa}\,,\qquad
\left\{\thetatb_0,\thetat_0\right\}=-\frac{1}{8\pi\kappa}\,,
\ee
with all other anti-commutators equal to zero. With the convention that $\xi_n$, $\xitb_n$, $\xi_0$ and $\thetat_0$
are the annihilaiton operators the normal ordered expression for $H_\kappa$ in the quantum theory is
\be
H_\kappa=2\kappa(\xib_0\xi_0+\thetatb_0\thetat_0)+
4\sum_{n\neq 0}\omega_n^2\left(\frac{\xib_n\xi_n}{\omega_n+\kappa}+\frac{\xit_n\xitb_n}{\omega_n-\kappa}\right)
+a_\kappa\,.
\ee
The normal ordering constant $a_\kappa$ is
\be
a_\kappa=\frac{1}{4\pi\kappa}\sum_{n=-\infty}^\infty\omega_n\,.\label{normordconst}
\ee
The remaining non-trivial bosonic N\"other current for the rotations
\be
\eta_1\rightarrow e^{i\epsilon}\eta_1\,,\qquad
\eta_2\rightarrow e^{-i\epsilon}\eta_2\,,
\ee
is
\be
j^c_\mu=\p_\mu\phi_+(\eta_1\eta^1-\eta_2\eta^2)+2i\eta_\mu{}^\nu\p_\nu\phi_+(\eta_2\eta_1+\eta^2\eta^1)\,.
\ee
The corresponding normal-ordered conserved current is
\ba
J&=&-\frac{1}{2\pi}\int d\sigma j^c_0=\frac{\kappa}{\pi}\int d\sigma(\eta_2\eta^2-\eta_1\eta^1)\nonumber \\
&=&2\kappa\thetatb_0\thetat_0-2\kappa\xib_0\xi_0+\sum_{n\neq 0}\omega_n
\left(\frac{\xit_n\xitb_n}{\omega_n-\kappa}-\frac{\xib_n\xi_n}{\omega_n+\kappa}\right)\,.
\ea
In this case the normal ordering constant is zero. Since $\phi_-$ is periodic in $\sigma$ we require that 
\be
0=\int^{2\pi}_0d\sigma \p_1\phi_-=i\int^{2\pi}_0d\sigma \eta_i\overleftrightarrow{\p_1}\eta^i\,,
\ee
In the quantum theory this is equivalent to the level matching requirement
\be
0=\sum_{n\neq 0}n\omega_n
\left(\frac{\xit_n\xitb_n}{\omega_n-\kappa}-\frac{\xib_n\xi_n}{\omega_n+\kappa}\right)
\left|\mbox{ physical }\right>\,.
\ee
Finally, we may compute the four non-zero supercharges
\ba
Q_1&\equiv& i\int_0^{2\pi}\frac{d\sigma}{2\pi}Q^1_{0\,4}=\kappa e^{-i\kappa\tau}\frac{d\sigma}{2\pi}\eta^1
=\kappa\thetatb_0\,,\\
Q_2&\equiv& i\int_0^{2\pi}\frac{d\sigma}{2\pi}Q^2_{0\,3}=\kappa e^{-i\kappa\tau}\frac{d\sigma}{2\pi}\eta^2
=\kappa\xib_0\,,\\
{\bar Q}_1&\equiv& i\int_0^{2\pi}\frac{d\sigma}{2\pi}Q^4_{0\,1}=\kappa e^{i\kappa\tau}\frac{d\sigma}{2\pi}\eta_1
=\kappa\thetat_0\,,\\
{\bar Q}_2&\equiv& i\int_0^{2\pi}\frac{d\sigma}{2\pi}Q^3_{0\,2}=\kappa e^{i\kappa\tau}\frac{d\sigma}{2\pi}\eta_2
=\kappa\xi_0\,,
\ea

\noindent The above (super-)charges form a $U(1|1)^2$ algebra and in particular we find
\be
\left[H_\kappa\,,J\right]=0\,,\qquad \left\{Q_i,{\bar Q}_j\right\}=-\frac{\kappa}{8\pi}\delta{ij}\,.
\ee

\section{Comments on conformal invariance of the action~(\ref{gsu112})}\label{appb}
\setcounter{equation}{0}

In this appendix we entertain the possibility of using the action~(\ref{gsu112}),~(\ref{gsu112comp}) 
as a Polyakov string action. As a warm-up let us integrate out $\phi_-$ in the action~(\ref{gsu112comp}). 
We arrive at an effective action for the fermions in which we may set
\be
\phi_+=2\kappa\tau\,.
\ee
Explicitly we then have
\be
{\cal L}_{\mbox{\scriptsize eff}}=(-\kappa)\int d^2\sigma\, 
i\eta_i\overleftrightarrow{\p_0}\eta^i-2\kappa\eta^i\eta_i
+\eta_1\overleftrightarrow{\p_1}\eta_2-\eta^1\overleftrightarrow{\p_1}\eta^2\,.
\ee
We may represent the  worldsheet gamma matrices as
\be
\rho^0=\left(\begin{array}{rr}-1&0\\ 0&1\end{array}\right)\,,\qquad
\rho^1=\left(\begin{array}{rr}0&-i\\ -i&0\end{array}\right)\,,
\ee
and define a world-sheet Dirac spinor as
\be
\psi_\a=\left(\begin{array}{c}\eta_1 \\ \eta^2\end{array}\right)\,,\qquad \a=1,2\,.
\ee
The conjugate spinor is then
\be
{\bar\psi}_\a=(\psi^\dg\rho^0)_\a=\left(-\eta^1, \eta_2\right)_\a\,,
\ee
and the effective action may be written as
\be
{\cal L}_{\mbox{\scriptsize eff}}=(-\kappa)\int d^2\sigma\, i\delta_a^\mu\psib\rho^a\overleftrightarrow{\p_\mu}\psi
+2\kappa\psib\psi
\,,
\ee
where
\be
\psib\rho^a\overleftrightarrow{\p_\mu}\psi\equiv
\psib\rho^a\p_\mu\psi-\p_\mu\psib\gamma^a\psi\,.
\ee
This is simply the Lagrangian for a worldsheet Dirac fermion of mass $2\kappa$. Since such fermions are not
conformal, we get the first indication that the Lagrangian~(\ref{gsu112}) is also not conformal.
                                                                                                                           
\noindent With the above definitions for $\rho^a$, $\psi_\a$ and $\psib_\a$ we can re-write the
action~(\ref{gsu112}) as
\be
{\cal L}=\int d^2\sigma\sqrt{-g}\left(g^{\mu\nu}\p_\mu\phi_+\p_\nu\phi_-+
ie^\mu_a\psib\rho^a\overleftrightarrow{\p_\mu}\psi
+2m\psib\psi\right)\,,\label{firstcurved}
\ee
where
%More generally we may
%not wish to set $\phi_+=2\kappa\tau$.  Integrating out $\phi_-$ fixes $\phi_+$ to satisfy
%\be
%\Box\phi_+ = 0\,.
%\ee
%Let us define an inverse zwei-bein
\be
e^\mu_a=\left(g^{\mu\nu}\p_\nu\phi_+\,,\frac{1}{\sqrt{-g}}\epsilon^{\mu\nu}\p_\nu\phi_+\right)\,.
\ee
We have written the above expression in the form of an inverse zwei-bein; we will see shortly that this
is indeed justified. The corresponding zwei-bein is
\be
e^a_\mu=\left(\p_\mu\phi_+\,,\epsilon_{\mu\nu}\p^\nu\phi_+\right)\,,
\ee
and the metric is
\be
G_{\mu\nu}\equiv e^a_\mu e^b_\nu\eta_{ab}=-\frac{1}{m}g_{\mu\nu}\,.
%m^{-2}\left(\begin{array}{cc}
%g(\p^1\phi_+)^2-(\p_0\phi_+)^2 & -g(\p^0\phi_+\p^1\phi_+) -(\p_0\phi_+\p_1\phi_+) \\
%-g(\p^0\phi_+\p^1\phi_+) -(\p_0\phi_+\p_1\phi_+) & g(\p^0\phi_+)^2-(\p_1\phi_+)^2\end{array}\right)\,.
\ee
Above
\be
m\equiv g^{\mu\nu}\p_\mu\phi_+\p_\nu\phi_+=\sqrt{-G^{\mu\nu}\p_\mu\phi_+\p_\nu\phi_+}\,,
\ee
is the norm of $\phi_+$, which needs to be non-zero. For completness note that
the determinant of the metric and the zwei-bein are
\be
G\equiv\det G_{\mu\nu} = \frac{g}{m^2}\,,\qquad e\equiv\det(e^a_\mu)=-\frac{\sqrt{-g}}{m}\,.
\ee
Rescaling fermions in the action~(\ref{firstcurved}) by
\be
\psi\rightarrow m^{-1/2}\psi\,,
\ee
gives
\be
{\cal L}=\int d^2\sigma\sqrt{-g}\left(g^{\mu\nu}\p_\mu\phi_+\p_\nu\phi_-+
im^{-1}e^\mu_a\psib\rho^a\overleftrightarrow{\p_\mu}\psi
+2\psib\psi\right)\,.
\ee
Integrating by parts this can be written as
\ba
{\cal L}&=&\int d^2\sigma\sqrt{-g}\left(g^{\mu\nu}\p_\mu\phi_+\p_\nu\phi_-+
2im^{-1}e^\mu_a
\psib\rho^a\p_\mu\psi+\frac{\p_\mu(e^\mu_a\sqrt{-g}m^{-1})}{\sqrt{-g}}\psib\rho^a\psi
+2\psib\psi\right)\nonumber \\
&=&\int d^2\sigma\sqrt{-g}\left(g^{\mu\nu}\p_\mu\phi_+\p_\nu\phi_-+
2im^{-1}e^\mu_a
\psib\rho^a\p_\mu\psi+m^{-1}\frac{\p_\mu(e^\mu_a\sqrt{G})}{\sqrt{G}}\psib\rho^a\psi
+2\psib\psi\right)\nonumber \\
&=&\int d^2\sigma\sqrt{-g}\left(g^{\mu\nu}\p_\mu\phi_+\p_\nu\phi_-+
2im^{-1}e^\mu_a
\psib\rho^a\p_\mu\psi+m^{-1}\omega_a^{01}\psib\rho^a\rho_{01}\psi
+2\psib\psi\right)\nonumber \\
&=&\int d^2\sigma\sqrt{-G}\left(-G^{\mu\nu}\p_\mu\phi_+\p_\nu\phi_-+
2\psib(i\rho^\mu D_\mu+m)\psi\right)
\,.\label{curved}
\ea
The final form of the action is that of a world-sheet Dirac fermion of mass $m$ together with the fields $\phi_\pm$
moving in a curved metric $G_{\mu\nu}$. Above we have used the fact that in two dimensions for any
zwei-bein ${\hat e}^a_\mu$ and corresponding metric ${\hat g}_{\mu\nu}$, the spin
connection ${\hat\omega}_\mu^{ab}$ can be written as
\be
{\hat \omega}_\mu^{ab}=-\epsilon^{ab}\frac{1}{\sqrt{{\hat g}}}{\hat e}^c_\mu\epsilon_c{}^d
\p_\nu\left({\hat e}^\nu_d\sqrt{{\hat g}}\right)\,,
\ee
where $\epsilon^{ab}$ ($\epsilon_c{}^d$) is the flat Minkowski space $\epsilon$-tensor with non-zero components
$\epsilon^{01}=-\epsilon^{10}=1$ ($\epsilon_0{}^1=\epsilon_1{}^0=-1$). This formula can be derived from the 
xpressions presentd in Appendix~\ref{appc}.
                                                                                                                           
\noindent
We may now want to define a string theory path integral for this Lagrangian. To do so we consider
the Polyakov path integral for the Lagrangian~(\ref{curved}). Since the path integral integrates over metrics
$g_{\mu\nu}$, and the Lagrangian is a function of the metric $G_{\mu\nu}=-m^{-1}g_{\mu\nu}$ we first rescale
\be
g_{\mu\nu}\rightarrow -m^{1/2} g_{\mu\nu}\,,
\ee
in order to eliminate the metric $G_{\mu\nu}$. We arrive at a Polyakov-type path-integral with action
\be
{\cal L}\rightarrow\int d^2\sigma\sqrt{-g}\left(g^{\mu\nu}\p_\mu\phi_+\p_\nu\phi_-+
2\psib(i\rho^\mu D_\mu+\sqrt{m})\psi\right)\,.\label{altu11}
\ee
This Lagrangian is conformally invariant. One way to see this is to generalise the argument presented 
in~\cite{prt} which considered sigma-models on plane-wave backgrounds. Let us integrate out the fermions to 
obtain an effective Lagrangian for $\phi_\pm$~\footnote{I am grateful to A. Tseytlin for a number of 
discussions and explanations of these issues.}
\ba
{\cal L}_{\mbox{\scriptsize eff}}&\sim&\eta^{\mu\nu}\p_\mu\phi_+\p_\mu\phi_-+
\frac{i}{2}\log\det\left[-\frac{\delta^2 {\cal L}}{\delta\eta\delta\eta}\right]
\nonumber \\
&=&\eta^{\mu\nu}\p_\mu\phi_+\p_\mu\phi_-+
\frac{i}{2}\log\det\left[\p_{\mu_1}\phi_+\Pi^{\mu_1\mu_2}_+\p_{\mu_2}\p_{\nu_1}\phi_+\Pi^{\nu_1\nu_2}_-\p_{\nu_2}
+\frac{1}{4}m^2\right]\nonumber \\
&=&\eta^{\mu\nu}\p_\mu\phi_+\p_\mu\phi_-
+\frac{i}{2}\det\left[\p^2+\frac{1}{4}m^2\right]
+\frac{i}{2}\log (m^2)\nonumber \\
&\sim&\eta^{\mu\nu}\p_\mu\phi_+\p_\mu\phi_-
+\eta^{\mu\nu}\p_\mu\phi_+\p_\mu\phi_+\ln\Lambda
\,,
\ea
where $\Lambda$ is the cut-off. We can re-absorb this divergent piece by re-defining $\phi_-$
\be
\phi_-\rightarrow\phi_--\phi_+\ln\Lambda\,.
\ee
This shows that the Lagrangian~(\ref{altu11}) is conformal. As it stands however,
this Lagrangian is not Weyl invariant and, just as in~\cite{prt}, we need to turn on a dilaton
\be
\Phi=\phi_+^2\,.
\ee

\section{Two dimensional spin connection}\label{appc}
\setcounter{equation}{0}

Let us consider a geenral Lorenzian two dimensional metric $g_{\mu\nu}$ which we will parametrise
for convenience as
\be
g_{\mu\nu}=\left(\begin{array}{cc} a^2& b \\ b& d^2 \end{array}\right)\,,
\ee
where $a$, $b$ and $d$ are complex functions of $\tau$ and $\sigma$ the coordinates on the manifold. The zwei-bein from which this follows is given
by
\be
e^1_\mu=(a\sinh\rho \,,-d\sinh\rho )\,,\qquad
e^2_\mu=(a\cosh\rho \,,d\cosh\rho )\,,
\ee
where
\be
\cosh\frac{1}{2}\rho=\frac{b}{ad}\,.
\ee
The Christoffel symbols 
\be
\Gamma^\mu_{\nu\lambda}=\frac{1}{2}g^{\mu\kappa}\left(g_{\kappa\nu\,,\lambda}
+g_{\kappa\lambda\,,\nu}-g_{\nu\lambda\,,\kappa}\right)
\ee
are given by
\ba
\Gamma^1_{11}&=&g^{-1}\left(aba_{,1}+ad^2a_{,0}-bb_{,0}\right)\,,\\
\Gamma^1_{12}&=&\Gamma^1_{21}=g^{-1}\left(d^2aa_{,1}-bdd_{,0}\right)\,,\\
\Gamma^1_{22}&=&g^{-1}\left(d^2b_{,1}-bdd_{,1}-d^3d_{,0}\right)\,,\\
\Gamma^2_{11}&=&g^{-1}\left(-a^3a_{,1}-aba_{,0}+a^2b_{,0}\right)\,,\\
\Gamma^2_{12}&=&\Gamma^1_{21}=g^{-1}\left(-aba_{,1}+a^2dd_{,0}\right)\,,\\
\Gamma^2_{22}&=&g^{-1}\left(a^2dd_{,1}-bb_{,1}+bdd_{,0}\right)\,.
\ea
where $g=\det g_{\mu\nu}$. It is easy to check that these satisfy the defining equation
\be
g_{\mu\nu\,,\lambda}-g_{\kappa\nu}\Gamma^\kappa_{\mu\lambda}
-g_{\kappa\mu}\Gamma^\kappa_{\nu\lambda}=0\,.
\ee
The spin connection $\omega^{mn}_\mu$ can be determined from the following equation
\be
D_\mu e^m_\nu=\p_\mu e^m_\nu+\omega_\mu^m{}_ne^n_\nu-\Gamma^\kappa_{\mu\nu}e^m_\kappa=0\,.
\ee
Since $\omega^{mn}_\mu$ is anti-symmetric in $(m,n)$ the non-zero components are given by
\ba
\omega^{01}_0&=&-\omega^{10}_0=
\frac{-2a^2da_{,1}-bda_{,0}+adb_{,0}+abd_{,0}}{2ad\sqrt{-g}}\,,\\
\omega^{01}_1&=&-\omega^{10}_1=\frac{-bda_{,1}-adb_{,1}+abd_{,1}+2ad^2d_{,0}}
{2ad\sqrt{-g}}\,.
\ea

\section{T-dual version of the action~(\ref{gsu112})}\label{appd}
\setcounter{equation}{0}

Performing T-duality for the action~(\ref{gsu112}) along $\a$ leads to a very simple form for an equivalent 
action. In this appendix we breifly present these results. To T-dualise along $\a$ we replace $\p_\mu\a$ by 
$A_\a$ and adding the Lagrange multiplier term $\epsilon^{\mu\nu}A_\mu\p_\nu\at$. The $A_\mu$ are then integrated 
out and we obtain the action
\ba
{\cal L}^d_\kappa&=&\int d^2\sigma\frac{\sqrt{g}g^{\mu\nu}}{1-\eta^i\eta_i}
\left(-(\p_\mu t-\frac{i}{2}\eta_i\overleftrightarrow{\p_\mu}\eta^i)
(\p_\nu t-\frac{i}{2}\eta_i\overleftrightarrow{\p_\nu}\eta^i)\right.
\nonumber \\&&\qquad\qquad\qquad\,\,\,\,\,\,\,\left.
+(\p_\mu\at-(\eta_1\overleftrightarrow{\p_\mu}\eta_2-\eta^1\overleftrightarrow{\p_\mu}\eta^2))
(\p_\nu\at-(\eta_1\overleftrightarrow{\p_\nu}\eta_2-\eta^1\overleftrightarrow{\p_\nu}\eta^2))
\right)
\nonumber \\&&\qquad\,\,\,\,\,\,\,+
\frac{2\epsilon^{\mu\nu}}{1-\eta^i\eta_i}
\left((\p_\mu t-\frac{i}{2}\eta_i\overleftrightarrow{\p_\mu}\eta^i)
(\p_\nu\at-(\eta_1\overleftrightarrow{\p_\nu}\eta_2-\eta^1\overleftrightarrow{\p_\nu}\eta^2))
\right)\label{tdual}
\ea
where we have used the fact that up to total derivatives
\be
\int d^2\sigma \frac{\epsilon^{\mu\nu}}{1-\eta^i\eta_i}\p_\mu t\p_\nu\at\eta^i\eta_i=
\int d^2\sigma \frac{\epsilon^{\mu\nu}}{1-\eta^i\eta_i}\p_\mu t\p_\nu\at\,.
\ee
At the level of classical equations of motion we may integrate out the metric to get a Nambu-Goto type action
\ba
{\cal L}^{d\, NG}_\kappa&=&\int d^2\sigma
\frac{\epsilon^{\mu\nu}}{1-\eta^i\eta_i}
\left((\p_\mu t-i\eta_i\overleftrightarrow{\p_\mu}\eta^i)
(\p_\nu\at-(\eta_1\overleftrightarrow{\p_\nu}\eta_2-\eta^1\overleftrightarrow{\p_\nu}\eta^2))
\right)\,,
\ea
where we have rescaled $t\rightarrow t/2$ and multiplied the whole action by a factor of 2.
The Nambu-Goto form of the action is particularily simple due to the 'two-dimensional' target space
form of the action~(\ref{tdual}). The equations of motion for $\at$ and $t$ imply that
\ba
\frac{1}{1-\eta^i\eta_i}(\p_\mu t-\frac{i}{2}\eta_i\overleftrightarrow{\p_\mu}\eta^i)&=&\p_\mu\chi_1\,,\\
\frac{1}{1-\eta^i\eta_i}(\p_\nu\at-(\eta_1\overleftrightarrow{\p_\nu}\eta_2-\eta^1\overleftrightarrow{\p_\nu}\eta^2)
&=&\p_\mu\chi_2\,,
\ea
where $\chi_i$ are arbitrary Grassmann-even functions of $\tau$ and $\sigma$.
The fermion equations of motion can then be written in form notation as
\be
0=\eta_i \mbox{d}\chi_1{}_\wedge \mbox{d}\chi_2-i\mbox{d}\eta_i{}_\wedge\mbox{d}\chi_2
+i\mbox{d}\chi_1{}_\wedge\mbox{d}\eta^j\,,
\ee
where $i\neq j$. 
                                                                                                                           
\noindent
Let us combine the ($1+1$ dimensional) spacetime coordinates into a two-vector
\be
x^i=(t,\at)^i\,,\qquad i=1,2\,,
\ee
and represent the  spacetime gamma matrices as
\be
\gamma^0=\left(\begin{array}{rr}-1&0\\ 0&1\end{array}\right)\,,\qquad
\gamma^1=\left(\begin{array}{rr}0&-i\\ -i&0\end{array}\right)\,,
\ee
a spacetime Dirac spinor as
\be
\Psi_\a=\left(\begin{array}{c}\eta_1 \\ \eta^2\end{array}\right)\,,\qquad \a=1,2\,.
\ee
The conjugate spinor is then
\be
{\bar\Psi}_\a=(\Psi^\dg\gamma^0)_\a=\left(-\eta^1, \eta_2\right)_\a\,.
\ee
With these definitions the Nambu-Goto action can be written as
\ba
{\cal L}^{d\, NG}_\kappa&=&\frac{1}{2}\int d^2\sigma
\frac{\epsilon_{ij}\epsilon^{\mu\nu}}{1+\Psi^2}
\Pi^i_\mu\Pi^j_\nu\,,
\ea
where we define
\be
\Pi^i_\mu\equiv(\p_\mu x^i-i\Psib\gamma^i\overleftrightarrow{\p_\mu}\Psi)\,,\qquad\Psi^2=\Psib\Psi\,,
\ee
and
\be
\Psib\gamma^i\overleftrightarrow{\p_\mu}\Psi\equiv
\Psib\gamma^i\p_\mu\Psi-\p_\mu\Psib\gamma^i\Psi\,.
\ee
%It is tempting to define the usual spacetime supersymmetry transformations
%\be
%\delta\Psi=\epsilon\,,\qquad
%\delta\Psib={\bar\epsilon}\,,\qquad
%\delta x^i=i\left({\bar\epsilon}\gamma^i\Psi+\Psib\gamma^i\epsilon\right)\,,
%\ee
%which would leave the action invariant were it not for the prefactor $1/(1+\Psib\Psi)$. Instead, if we modify the
%variation of $x^i$ such that
%\be
%\delta \p_\mu x^i=i\left({\bar\epsilon}\gamma^i\Psi+\Psib\gamma^i\epsilon\right)
%-\frac{{\bar\epsilon}\Psi+\Psib\epsilon}{1+\Psi^2}\Pi_\mu^i\,,
%\ee
%we do obtain a fermionic symmetry of the action.

\end{document}